\documentclass[%
 reprint,
superscriptaddress,
nofootinbib,
 amsmath,amssymb,
 aps,
]{revtex4-2}

\usepackage{graphicx}
\usepackage{dcolumn}
\usepackage{bm}
\usepackage{hyperref}
\usepackage[usenames,dvipsnames]{xcolor}
\hypersetup{
    colorlinks = true,
    citecolor = {MidnightBlue},
    linkcolor = {BrickRed},
    urlcolor = {BrickRed}
}

\usepackage{aas_macros}

\newcommand{\F}{\mathcal{F}}	
\newcommand{\G}{\mathcal{G}}	
\newcommand{\Pow}{\mathcal{P}}	

\begin{document}

\preprint{APS/123-QED}

\title{Cosmic flexion}

\author{Evan J. Arena}
  \email{evan.james.arena@drexel.edu}
  \homepage{http://evanjarena.github.io}
 \affiliation{%
Department of Physics, Drexel University, Philadelphia, PA 19104, USA
}%
\author{David M. Goldberg}%
\affiliation{%
Department of Physics, Drexel University, Philadelphia, PA 19104, USA
}%
\author{David J. Bacon}
\affiliation{
Institute of Cosmology and Gravitation, University of Portsmouth, Burnaby Road, Portsmouth PO1 3FX, UK\
}%

\date{\today}

\begin{abstract}
Cosmic flexion, like cosmic shear, is a correlation function whose signal originates from the large-scale structure of the Universe.  Building on the observational success of cosmic shear, along with the unprecedented quality of large-scale cosmological datasets, the time is ripe to explore the practical constraints from cosmic flexion. Unlike cosmic shear, which has a broad window function for power, cosmic flexion is only measurable on small scales and therefore can uniquely place constraints on the small-scale matter power spectrum.  Here, we present a full theoretical formalism for cosmic flexion, including both flexion-flexion and shear-flexion two-point correlations. We present forecasts for measuring cosmic flexion in the Dark Energy Survey (DES), a Stage III cosmological survey, and comment on the future prospects of measuring these cosmological flexion signals in the upcoming era of Stage IV experiments.
\end{abstract}

\maketitle


\section{\label{sec:intro}Introduction}

Cosmological studies of the Cosmic Microwave Background (CMB) have found that observations agree with the standard model of cosmology ($\Lambda$CDM) to remarkable accuracy \citep{Blumenthal:1984bp, SupernovaSearchTeam:1998fmf, SupernovaCosmologyProject:1998vns, Aghanim:2018eyx}. As we look at more recent parts of cosmic history, using tools such as weak lensing, $\Lambda$CDM still appears to be the law of the land. Subtle discrepancies are found, however,  between these low-redshift observations and the high-redshift measurements of the CMB \cite{DES:2021wwk, Aghanim:2018eyx, Riess:2020fzl, Huang:2019yhh}. These discrepancies could indicate that $\Lambda$CDM might not be sufficient to connect all parts of the cosmic history \citep{Frusciante:2019xia, Bloomfield:2012ff, Bellini:2014fua}.  It is therefore necessary to have multiple cosmological probes that complement each other in order to get the full picture of cosmology across all length scales and cosmic times.    

Gravitational lensing has become one of the quintessential cosmological and astrophysical probes of the last few decades \citep{Bartelmann:1999yn, Kilbinger:2014cea, DES:2021vln}. Lensing probes the gravitational potential and is therefore a useful measure of the total matter distribution.  To this end, lensing has had a great impact at several different mass scales. Lensing is powerful for studying galaxy cluster mass distributions \citep{1987A&A...172L..14S, 1990ApJ...349L...1T}.  A weaker effect, known as galaxy-galaxy lensing, is the lensing of a background galaxy by a foreground galaxy. Specifically, galaxy-galaxy shear correlates the shapes of high-redshift ``source'' galaxies with positions of low-redshift ``lensing'' galaxies \citep{Brainerd:1995da, SDSS:1999zww}.  Even weaker is the lensing by the large-scale structure of the Universe -- specifically, the so-called cosmic shear -- which probes the underlying matter power spectrum \citep{Bacon:2000sy, Kaiser:2000if, vanWaerbeke:2000rm, Wittman:2000tc}. Finally, lensing of the CMB has also been detected at high significance \cite{Aghanim:2018eyx}, which has been a further useful probe of cosmological parameters. 

In studies interested in using low-redshift lensing measurements to extract cosmological information, cosmic shear is often combined with galaxy-galaxy lensing, along with galaxy clustering, which allows for very high constraining power on cosmological parameters \cite{DES:2021wwk}.

Beyond shear, there exists a higher-order lensing effect known as \textit{flexion} \citep{Goldberg:2004hh, Bacon:2005qr, Okura:2007js, Schneider:2007ks}.  In this work, we will consider the as-of-yet undetected cosmic flexion signal, the analogue to cosmic shear.  Cosmic flexion is the flexion correlation function whose signal originates from the large-scale structure of the Universe.  Cosmic shear has proven to be a highly valuable cosmological probe; cosmic flexion therefore warrants further investigation in order to determine the extent to which it is cosmologically useful.  Cosmic flexion is much more difficult to detect than cosmic shear, owing to (i) its weaker signal-to-noise ratio (S/N) on the scale of typical galaxy-galaxy separation, (ii) the fact that it peaks at small, nonlinear scales, and (iii) the lack of an appropriate tool for measuring this signal -- until now.  We also consider cosmic shear-flexion -- i.e., the cross-correlation between cosmic shear and flexion -- which has a higher S/N than flexion-flexion, albeit at different angular scales.   

We will first present the theory of cosmic flexion. We then consider the feasibility of detecting this signal in Stage III lensing surveys such as the Dark Energy Survey\footnote{\href{https://www.darkenergysurvey.org}{https://www.darkenergysurvey.org}} (DES; \cite{DES:2016jjg}), the Kilo-Degree Survey\footnote{\href{https://kids.strw.leidenuniv.nl}{https://kids.strw.leidenuniv.nl}} (KiDS; \cite{Kuijken:2015vca}) and the Hyper Suprime-Cam Subaru Strategic Program\footnote{\href{https://hsc.mtk.nao.ac.jp/ssp}{https://hsc.mtk.nao.ac.jp/ssp}} (HSC SSP; \cite{Aihara2017TheHS}), with the aim of constraining the normalization and slope of small-scale cosmic structure.  We then comment on improvements from Stage IV surveys such as the Legacy Survey of Space and Time with the Vera C. Rubin Observatory\footnote{\href{https://www.lsst.org}{https://www.lsst.org}} (LSST; \cite{LSSTScience:2009jmu}), Euclid\footnote{\href{https://www.euclid-ec.org}{https://www.euclid-ec.org}} \cite{Refregier:2010ss}, and the Nancy Grace Roman Space Telescope\footnote{\href{https://roman.gsfc.nasa.gov}{https://roman.gsfc.nasa.gov}} \cite{2015arXiv150303757S}, as well as physics that may be constrained by these results.

\section{Theory of Cosmic Flexion}

\subsection{Lensing Formalism}
\label{sec:lensing_formalism} 

In the thin lens approximation, we relate the convergence, $\kappa$, to a dimensionless lensing potential, $\psi$, with $\nabla^2 \psi = 2\kappa$. The lensing potential is the two-dimensional analogue of the Newtonian gravitational potential, integrated along the line of sight. Convergence is a key lensing quantity, which can be simultaneously thought of as a projected, dimensionless surface-mass density of matter, and as an isotropic increase or decrease of the observed size of a source image.   In the weak lensing regime, the coordinate mapping from the foreground, ${\bm \theta}$, to background, ${\bm \beta}$, positions is related to the potential via: $\beta_i = \delta_{ij}\theta^j - \psi,_{ij}\theta^j - \frac{1}{2}\psi,_{ijk}\theta^j\theta^k$ (where $\psi,_{ij}$ is shorthand for $\partial_i\partial_j\psi$).  We define a complex gradient operator $\partial = \partial_1+i\partial_2$, such that 1 and 2 refer to two perpendicular directions locally on the sky (i.e. $x$ and $y$ directions on an image of a small patch of sky).  In this formalism, the spin-2 shear is given by \cite{Bacon:2005qr}
\begin{equation}
    \gamma = \gamma_1 + i\gamma_2 = |\gamma|e^{2i\phi} = \frac{1}{2}\partial\partial\psi
\end{equation}

\noindent and the spin-1 and spin-3 flexion fields are given by the derivatives of the convergence and shear, respectively:
\begin{align}
    \F &= \F_1 + i\F_2 = |\F|e^{i\phi}  = \frac{1}{2}\partial\partial^*\partial\psi = \partial\kappa, \\
    \G &= \G_1 + i\G_2 = |\G|e^{3i\phi} =\frac{1}{2}\partial\partial\partial\psi = \partial\gamma. 
    \label{eq:FG}
\end{align}

\noindent The shear is an anisotropic, elliptical stretching of the source image.  The $\F$-flexion effect is a skewing distortion which manifests as a centroid shift, whereas the $\G$-flexion is a trefoil distortion resulting in a triangularization of the source image.

\subsection{Cosmic Lensing Power Spectra}
\label{sec:power_spectra} 

Starting with the cosmological effective convergence as in Ref. \cite{Bartelmann:1999yn}, one can obtain the convergence power spectrum via Limber's equation \citep{1953ApJ...117..134L, LoVerde:2008re}:
\begin{equation}\label{eqn:P_kappa}
    \Pow_{\kappa}(\ell) = \int_0^{\chi_H} d\chi \frac{q^2(\chi)}{\chi^2}\Pow_{\rm NL}\left(k = \frac{\ell+1/2}{\chi},\chi \right)
\end{equation}

\noindent where the lensing efficiency function 
\begin{equation}\label{eqn:q}
q(\chi) = \frac{3}{2}\Omega_m \left(\frac{H_0}{c}\right)^2\frac{\chi}{a(\chi)} \int_{\chi}^{\chi_H}d\chi' n(\chi') \frac{\chi' - \chi}{\chi'}.
\end{equation}

\noindent In these expressions, $H_0$ is the Hubble constant, $\Omega_m$ is the matter density at the present epoch, $c$ is the speed of light, $\chi$ is comoving distance, $\chi_H$ is the horizon distance, $a$ is the scale factor, $k$ is the comoving wavenumber, $\Pow_{\rm NL}(k,z)$ is the (nonlinear) matter power spectrum\footnote{We caution the reader that the nonlinear matter power spectrum must be used, as the cosmic flexion signal exists in the small-scale, nonlinear regime.} as a function of $k$ and redshift, $z$, and $n(\chi)$ is the effective number density of (source) galaxies as a function of $\chi$, normalized such that $\int_0^\infty d\chi n(\chi) = 1$. The lensing efficiency function, and hence the power spectrum, depends on the redshift distribution of galaxies, $n(z)$, which is in turn dependent on the galaxies available for a particular cosmological survey.

In the case of cosmic shear, it is the shear that is measured from observed galaxy shapes, not the convergence.  However, it is conveniently the case that \cite{Kilbinger:2014cea}
\begin{equation}
    \Pow_\gamma(\ell) = \Pow_\kappa(\ell).
\end{equation}

\noindent In Ref. \cite{Bacon:2005qr} (hereafter referred to as BGRT), it was shown that a cosmic flexion power spectrum can be derived along the same lines, with the additional step of differentiating the cosmological effective convergence, and then making use of Limber's equation.  From this, we obtain
\begin{equation}\label{eqn:P_F}
    \Pow_\F(\ell) = \ell^2 \Pow_\kappa(\ell).
\end{equation}

\noindent We also note that, owing to the fact that the shear and convergence statistics are the same, so too (because of the relations in Eq. \eqref{eq:FG}) are the $\F$- and $\G$-flexion power spectra:
\begin{equation}
    \Pow_\G(\ell) = \Pow_\F(\ell).
\end{equation}

\noindent BGRT also introduced the idea of a convergence-flexion cross spectrum.  Again following Limber's equation, one finds
\begin{equation}\label{eqn:P_kappaF}
    \Pow_{\kappa\F}(\ell) = \Pow_{\kappa\G}(\ell) = \ell \Pow_\kappa(\ell).
\end{equation}
We will use these power spectra later in the calculation of measurable two-point correlation functions.

\subsection{Two-Point Correlation Functions: Cosmic Shear and Flexion}

While one can try to measure the cosmic flexion (or shear) power spectra defined in Fourier space, it is often more straightforward to take measurements in real space.  One can calculate real-space two-point correlation functions by taking a Hankel transform of the power spectrum.  BGRT did this; however, they considered what turns out to be only one out of six possible nonzero cosmic flexion correlation functions.

\subsubsection{Coordinate System}

The shear and flexion correlation functions are defined by considering pairs of positions of galaxy images on the sky, ${\bm \vartheta}$ and ${\bm \theta}+{\bm \vartheta}$, and defining a coordinate system along the line connecting the two galaxies \cite{Kilbinger:2014cea}. For shear, $\gamma=\gamma_1 + i\gamma_2$; these components are conventionally referred to as the ``tangential'' and ``cross'' components. These are defined at position ${\bm \vartheta}$ for this pair as $\gamma_{\rm t} = -\Re\{\gamma e^{-2i\varphi}\}$ and $\gamma_{\times} = -\Im\{\gamma e^{-2i\varphi}\}$, respectively, where  $\varphi$ is the polar angle of the separation vector $\bm{\theta}$.

This spin-2 cosmic shear coordinate system can be generalized to any combination of spin fields. Let $a = a_1 + ia_2$ and $b = b_1 + ib_2$ be two fields with spins $s_a$ and $s_b$.  Define $a'(\bm{\vartheta}_i)$ and $b'(\bm{\vartheta}_j)$ as the fields $a$ and $b$ at locations $\bm{\vartheta}_i$ and $\bm{\vartheta}_j$ rotated such that the $x$-axis of the tangential coordinate systems at directions $\hat{\bm{\vartheta}}_i$ and $\hat{\bm{\vartheta}}_j$ become aligned with the vector connecting both points. (Note: $\hat{\bm{\vartheta}}_i\cdot\hat{\bm{\vartheta}}_j = \cos\theta$ and $\bm{\theta} = \bm{\vartheta}_j - \bm{\vartheta}_i$.) We may then define the components in this rotated coordinate system as

\begin{equation}\label{eqn:SO(2)}
    \begin{pmatrix}a_1'\\a_2'\end{pmatrix} = {\rm csgn}\left[(-i)^{s_a}\right] R(s_a\varphi)\begin{pmatrix}a_1\\a_2\end{pmatrix}
\end{equation}

\noindent where ${\rm csgn}(z)$ is the complex signum function and the (passive) rotation matrix is defined as
\begin{equation}
    R(s_a\varphi) = \begin{pmatrix}\cos s_a\varphi & \sin s_a\varphi\\ -\sin s_a\varphi & \cos s_a\varphi\end{pmatrix}.
\end{equation}

\noindent We choose to adopt this SO(2) formalism rather than the conventional real- and imaginary-component formalism from the literature, as we believe it more straightforwardly demonstrates that this is a rotated coordinate system. We see, then, that
\begin{equation}\label{eqn:gam_t_cross}
   \begin{pmatrix}\gamma_1'\\\gamma_2'\end{pmatrix} = - R(2\varphi)\begin{pmatrix}\gamma_1\\\gamma_2\end{pmatrix}.
\end{equation}

\noindent In the same way, we can define the rotated components of the lensing flexions $\F=\F_1 + i\F_2$ and $\G=\G_1 + i\G_2$ as 
\begin{align}\label{eqn:FtFr}
    \begin{pmatrix}\F_1'\\\F_2'\end{pmatrix} &= - R(\varphi)\begin{pmatrix}\F_1\\\F_2\end{pmatrix} \\
    \label{eqn:GtGr}
    \begin{pmatrix}\G_1'\\\G_2'\end{pmatrix} &= + R(3\varphi)\begin{pmatrix}\G_1\\\G_2\end{pmatrix}.
\end{align}

\noindent It should be pointed out that, in this work, the conventional tangential and cross components of the shear, $(\gamma_{\rm t}, \gamma_\times)$, are referred to as $(\gamma_1', \gamma_2')$. The conventional names refer to the fact that $\gamma_1' > 0$ corresponds to tangential alignment of galaxies around an overdensity, and the cross-component is oriented along a $45^\circ$ angle with respect to the line connecting the galaxy pair.  With the spin-1 $\F$-flexion, however, there is radially inward alignment around an overdensity, such that a tangential $\F$-flexion is analogous to a cross shear.  To avoid the confusion arising from these different directional alignments of various spin fields, we instead choose to refer rather plainly to rotated 1- and 2-components.  Furthermore, ${\rm csgn}\left[(-i)^{s_a}\right]$ is introduced such that the $\G$-flexion has what can roughly be thought of as a radially outward alignment around an overdensity, as motivated by its behavior around a Singular Isothermal Sphere (SIS) lens (see BGRT, where $\gamma < 0$ and $\F < 0$ around an SIS lens, but $\G > 0$).

\subsubsection{Real-Space Two-Point Correlation Functions}

It is well known with cosmic shear that one can construct three two-point correlations from the two shear components, $\langle \gamma_1'\gamma_1'\rangle$, $\langle \gamma_2'\gamma_2'\rangle$, and $\langle \gamma_1'\gamma_2'\rangle$ \cite{Kilbinger:2014cea}.  The latter vanishes in a parity-symmetric Universe, since $\gamma_1'$ is parity invariant under a mirror transformation, but $\gamma_2'$ changes sign.  The two nonzero correlations are then combined into the two components of the cosmic shear correlation functions.

In general, we can define two correlation functions \cite{Chisari_2019}:
\begin{align}
    \label{eqn:2PCF_general_plus}
     \xi_+^{ab}(\theta) &= \Re\langle a'(\bm{\vartheta}_i)b'^*(\bm{\vartheta}_j)\rangle = \langle a_1'b_1' \rangle + \langle a_2'b_2' \rangle \\
     \label{eqn:2PCF_general_minus}
    \xi_-^{ab}(\theta) &= \Re\langle a'(\bm{\vartheta}_i) b'(\bm{\vartheta}_j)\rangle = \langle a_1'b_1' \rangle - \langle a_2'b_2' \rangle. 
\end{align}

\noindent where $1$ and $2$ refer to the components of each field and $\langle a_1'b_1' \rangle$ is shorthand for $\langle a_1'(\bm{\vartheta}_i)b_1'(\bm{\vartheta}_j) \rangle$. Therefore, in addition to the well known cosmic shear correlation functions\footnote{$\xi_{\pm}^{\gamma\gamma}$ is referred to simply as $\xi_{\pm}$ in the cosmic shear literature, owing to the fact that it is currently the only lensing field correlation that is widely considered.}
\begin{equation}
    \xi_{\pm}^{\gamma\gamma}(\theta) = \langle \gamma_1'\gamma_1'\rangle \pm \langle \gamma_2'\gamma_2'\rangle,
\end{equation}

\noindent we posit the existence of six cosmic flexion correlation functions.  Firstly, there are the autocorrelations of each flexion field
\begin{align}
    \xi_{\pm}^{\F\F}(\theta) &= \langle \F_1'\F_1' \rangle \pm \langle \F_2'\F_2' \rangle \\
    \xi_{\pm}^{\G\G}(\theta) &= \langle \G_1'\G_1' \rangle \pm \langle \G_2'\G_2' \rangle. 
\end{align}

\noindent Secondly, there is a cross-correlation between the two flexion fields (we will see that this is nonzero in Section \ref{sec:mixed_spin_consequences} below):
\begin{equation}
    \xi_{\pm}^{\F\G}(\theta) = \langle \F_1'\G_1' \rangle \pm \langle \F_2'\G_2' \rangle. 
\end{equation}

\noindent Of these six possible correlations, only $\xi_{+}^{\F\F}(\theta)$ was considered in BGRT. 

In addition to the shear-shear and flexion-flexion correlations listed above, we further posit the existence of shear-flexion cross-correlations, given by\footnote{One may be curious as to why we choose the ordering shear-flexion for $\gamma\F$, but flexion-shear for $\G\gamma$.  Simply put, we choose to have a convention where the spin of the first field is greater than or equal to that of the second.}  
\begin{align}
    \xi_{\pm}^{\gamma\F}(\theta) &= \langle \gamma_1'\F_1' \rangle \pm \langle \gamma_2' \F_2' \rangle \\
    \xi_{\pm}^{\G\gamma}(\theta) &= \langle \G_1'\gamma_1' \rangle \pm \langle \G_2'\gamma_2' \rangle.
\end{align}

\begin{figure}
  \centerline{\includegraphics[width=3in]{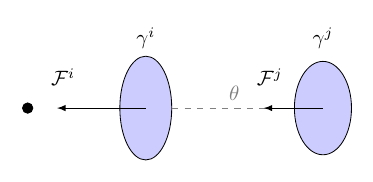}}
  \caption{A galaxy pair $(i,j)$ with separation $\theta$ and polar angle $\varphi_{ij}=0$ perturbed by a nearby mass distribution (on the left).  In this case, the overdensity is along the line of the separation vector such that we have pure tangential shear, $\gamma_1'$ and radial flexion, $\F_1'$ (we assume that there is no intrinsic ellipticity or flexion in this case).  The shear is represented by the ellipses and the spin-$1$ $\F$-flexion by the vectors.  As galaxy $(i)$ is closer to the overdensity, we see how the magnitude of the shear and flexion is larger for galaxy $(i)$ than $(j)$.  This cartoon illustrates how, for a given galaxy pair, flexion and shear are coupled between the
  objects, and to each other.}
\label{fg:diagram}
\end{figure}

\noindent Again, we will find these to be non-zero in Section \ref{sec:mixed_spin_consequences}. In Fig. \ref{fg:diagram}, we show a cartoon of the auto- and cross-correlations of $\F$-flexion and shear for a galaxy pair in real space.

\subsection{Relating the Correlation Functions to Power Spectra}

In the flat-sky approximation, the two-point correlation functions are related to the angular power spectra via \cite{Chisari_2019, Ng:1997ez, Chon:2003gx}
\begin{equation}\label{eqn:real_fourier_general}
    \xi_{\pm}^{ab}(\theta) = (\pm 1)^{s_a}\int_0^\infty \frac{d\ell\,\ell}{2\pi}\Pow_{ab}(\ell)J_{s_b \mp s_a}(\ell\theta)
\end{equation}

\noindent where $J_n(x)$ is the Bessel function of the first kind, order $n$. We do not derive this general equation in this paper.  Rather, it is a modified version of that presented in Ref. \cite{Chisari_2019}, where we have swapped $s_a$ and $s_b$ (we refer the reader to Appendix \ref{sec:appendix_A} for the justification of this).

From this general expression, we recover the well known relationship between the cosmic shear correlation functions and the convergence power spectrum:
\begin{equation}
    \xi_{\pm}^{\gamma\gamma}(\theta) = \int_0^\infty \frac{d\ell\,\ell}{2\pi}\Pow_{\kappa}(\ell)J_{0,4}(\ell\theta)
\end{equation}

\noindent where $J_0(\ell\theta)$ and $J_4(\ell\theta)$ refer to $\xi_{+}^{\gamma\gamma}$ and $\xi_{-}^{\gamma\gamma}$, respectively.  The flexion-flexion correlation functions are then given by
\begin{align}
    \label{eqn:xi_FF_theory}
    \xi_{\pm}^{\F\F}(\theta) &= \pm\int_0^\infty \frac{d\ell\,\ell}{2\pi}\Pow_{\F}(\ell)J_{0,2}(\ell\theta) \\
    \label{eqn:xi_GG_theory}
    \xi_{\pm}^{\G\G}(\theta) &= \pm\int_0^\infty \frac{d\ell\,\ell}{2\pi}\Pow_{\F}(\ell)J_{0,6}(\ell\theta) \\ 
    \label{eqn:xi_FG_theory}
        \xi_{\pm}^{\F\G}(\theta) &= \pm\int_0^\infty \frac{d\ell\,\ell}{2\pi}\Pow_{\F}(\ell)J_{2,4}(\ell\theta) 
\end{align}
 
\noindent and the shear-flexion correlation functions are

\begin{align}
    \label{eqn:xi_gammaF_theory}
    \xi_{\pm}^{\gamma\F}(\theta) &= \mp\int_0^\infty \frac{d\ell\,\ell}{2\pi}\Pow_{\kappa\F}(\ell)J_{1,3}(\ell\theta) \\
    \label{eqn:xi_Ggamma_theory}
    \xi_{\pm}^{\G\gamma}(\theta) &= -\int_0^\infty \frac{d\ell\,\ell}{2\pi}\Pow_{\kappa\F}(\ell)J_{1,5}(\ell\theta).  
\end{align}

\noindent In Appendix \ref{sec:appendix_A}, we show how to derive some of these correlation functions from first principles and demonstrate that they are in agreement with Eq. \eqref{eqn:real_fourier_general}.
 
\subsection{Consequences of Mixed Spin Field Cross-Correlation}\label{sec:mixed_spin_consequences}

The correlation of two different lensing fields is not widely considered in the literature. Only combinations of the same lensing field are generally discussed (i.e. shear-shear correlation).  Here, we discuss the implications of correlating lensing fields of different spin.  Note that throughout this discussion, ``spin combination" refers to the sum and/or difference of the spin fields of two correlated fields, $s_a\pm s_b$.  As such, spin combination can either be even, as in the case of cosmic shear or any other two-point autocorrelation, or odd.

\subsubsection{Consequence 1: Order Matters for Odd Spin Combinations, or the Noncommutativity of Weak Lensing}

Let us first consider cosmic shear.  One might intuitively guess that $\langle \gamma_1'\gamma_2' \rangle = 0$.  After all, the tangential and cross components are, by definition, not activated in the same way gravitationally.  Indeed, it turns out that $\langle \gamma_1'\gamma_2' \rangle$ vanishes due to the {\it parity symmetry} of the Universe.  Roughly speaking, if one were to look at the Universe under a mirror transformation, the combinations $\langle \gamma_1'\gamma_1' \rangle$ and $\langle \gamma_2'\gamma_2' \rangle$ would look the same (i.e. they are parity invariant), whereas $\langle \gamma_1'\gamma_2' \rangle$ would not.  The fact that $\langle \gamma_1'\gamma_2' \rangle$ is not parity invariant means that it must be zero in our parity-symmetric Universe. 

These arguments hold for cosmic flexion as well, for both $\F$ and $\G$.  One finds that $\langle \F_1'\F_1' \rangle$, $\langle \F_2'\F_2' \rangle$, $\langle \G_1'\G_1' \rangle$, $\langle \G_2'\G_2' \rangle$, $\langle \F_1'\G_1' \rangle$, $\langle \F_2'\G_2' \rangle$, etc., are parity-invariant combinations, whereas $\langle \F_1'\F_2' \rangle$, $\langle \G_1'\G_2' \rangle$,  $\langle \F_1'\G_2' \rangle$, etc., are not and will equal zero.

One might suppose that parity-symmetry requirements could pose a problem for the existence of a shear-flexion cross-correlation.  Consider a $\gamma$-$\F$ correlation.  There are four possible two-point correlations: $\langle \gamma_1'\F_1' \rangle$, $\langle \gamma_2'\F_2' \rangle$, $\langle \gamma_1'\F_2' \rangle$, and $\langle \gamma_2'\F_1' \rangle$.  We should immediately expect that  $\langle \gamma_1'\F_2' \rangle = \langle \gamma_2'\F_1' \rangle = 0$ due to parity symmetry.  This is indeed the case.  

However, it also could seem as though neither $\langle \gamma_1'\F_1' \rangle$ nor $\langle \gamma_2'\F_2' \rangle$ are parity invariant either. Recall that shear is spin-2, and $\F$ and $\G$ flexions are spin-1 and spin-3, respectively.  The spin-combinations are even for both shear-shear and flexion-flexion correlation.  Even spin implies a possible parity-invariant combination of components.  But any shear-flexion correlation will always have an odd-spin combination.  
This might appear to be an argument for any shear-flexion cross-correlation vanishing in our Universe. 

\begin{figure}
  \centerline{\includegraphics[width=\linewidth]{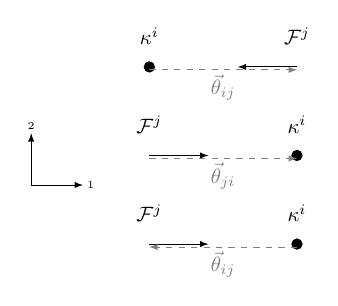}}
  \caption{Three different galaxy-galaxy flexion scenarios. In each scenario, the flexion has only a 1-component and points radially toward the overdensity shown by the convergence $\kappa^{i}$. From top to bottom, we refer to these as scenarios 1, 2, and 3. The separation vector $\vec{\theta}$ in scenarios 1 and 2 points from left to right, along the 1-axis, and from right to left in scenario 3. The polar angle $\varphi$ (i.e. the angle between the 1-axis and the separation vector) is $0$ radians for scenarios 1 and 2, and $-\pi$ radians for the scenario 3. Using Eq. \eqref{eqn:FtFr}, the flexion $\F_1'^j$ is $+|\F^{j}|$ for scenarios 1 and 3, and $-|\F^{j}|$ for scenario 2.}
\label{fg:gal-gal-F}
\end{figure}

In reality, certain odd-spin constructions do not simply vanish in this way.  In order to demonstrate this, consider first the example of galaxy-galaxy flexion (the flexion version of galaxy-galaxy shear).  In galaxy-galaxy shear, there is a tangential alignment of galaxy ellipticities around a foreground lens. In galaxy-galaxy flexion, there is a radial alignment of background galaxy flexions around the lens. Fig. \ref{fg:gal-gal-F} shows three different galaxy-galaxy flexion scenarios.  From top to bottom, let us refer to these as scenarios 1, 2, and 3, respectively. In scenario 1, there is an $\F$-flexion to the right of an overdensity, and in scenarios 2 and 3, the $\F$-flexion is to the left of the same overdensity. 

We notice that the flexion in scenario 2 is the negative of the flexion in scenario 1.  This sign difference might naively suggest that a galaxy-galaxy flexion signal vanishes (imagine adding these two flexions together), which we understand not to be the case -- galaxy-galaxy flexion has been measured in multiple scenarios (see e.g. Refs. \citep{Goldberg:2004hh,2007ApJ...666...51L,2011MNRAS.412.2665V}).  Conversely, the flexion in scenario 3 \textit{does} carry the same sign as that in scenario 1.  The difference between scenarios 2 and 3 is simply the direction of the separation vector, which is a result of the order in which the pairing is constructed. In scenarios 1 and 3, the pairing is $i\rightarrow j$, but in scenario 2, the pairing is $j\rightarrow i$.  The crucial point here is that the \textit{order} in which one field is rotated and correlated with another matters for odd-spin combinations, whereas all three of these scenarios give the same sign for an even spin combination such as galaxy-galaxy shear. We therefore distinguish between $\kappa\rightarrow\F$ and $\F\rightarrow \kappa$ correlation. $\kappa\rightarrow\F_k'$ denotes $\langle\kappa\F'_k\rangle$ where the separation vector $\vec\theta$ used for rotation points from a first object for which we supply $\kappa$, to a second object for which we supply $\F'_k$. $\F_k'\rightarrow\kappa$ is defined in a similar way, with $\vec\theta$ pointing from a first object for which we supply $\F'_k$ to a second object for which we supply $\kappa$. $\kappa\rightarrow\F_k'$ and $\F_k'\rightarrow\kappa$ turn out to be the negative, or the parity transforms, of each other.

This line of reasoning directly extends to shear-flexion cross-correlation.  With this odd spin combination, we need to emphasize the difference between $\gamma\rightarrow\F$ and $\F\rightarrow\gamma$ correlation.  Note that $\langle \gamma_1'\rightarrow\F_1' \rangle$ and $\langle \F_1'\rightarrow\gamma_1' \rangle$ are the parity transformations of each other, and carry opposite signs. The easiest way to visualize this is to recognize that $\gamma_1'$ is \textit{itself} parity invariant, whereas $\F_1'$ is not. As before, when one does a mirror transform of $\F_1'$, it is simply the negative of itself.  Hence
\begin{equation}\label{eqn:gammaF_parity_real}
    \langle \F_1'\rightarrow\gamma_1' \rangle = -\langle \gamma_1'\rightarrow\F_1' \rangle.
\end{equation}
This fact is also evident in the real-Fourier space relation.  From Eq. \eqref{eqn:real_fourier_general}, we see that
\begin{equation}\label{eqn:gammaF_parity_Fourier}
    \xi_{\pm}^{\F\rightarrow\gamma}(\theta) =  - \int_0^\infty \frac{d\ell\,\ell}{2\pi}\Pow_{\kappa\F}(\ell)J_{-1,3}(\ell\theta) = - \xi_{\pm}^{\gamma\rightarrow\F}(\theta)
\end{equation}

\noindent where we have used
\begin{equation}
    J_{-n}(x) = (-1)^n J_n(x).
\end{equation}

\noindent This is all to say that parity invariance for $\gamma-\F$ cross-correlation may be preserved through {\em fixed-ordered} pairing $i\rightarrow j$, and hence by distinguishing between $\gamma\rightarrow\F$ and $\F\rightarrow\gamma$.

As a final point, we note that {\em non-ordered} parity invariance is achieved by the fact that, while $\langle \F_1'\rightarrow\gamma_1' \rangle$ is nonzero (and hence measurable), it is indeed the case that the sum $\langle \F_1'\rightarrow\gamma_1' \rangle + \langle \gamma_1'\rightarrow\F_1' \rangle = 0$.

\subsubsection{Consequence 2: Mixed lensing field correlations provide information from more pairs}

One can measure both $\xi_{\pm}^{\F\rightarrow\G}$ and $\xi_{\pm}^{\G\rightarrow\F}$ for a given set of galaxy pairs (for instance, selected as in Eq. \eqref{eqn:Np} as galaxies $i,j>i$).  This is an example of how cross-correlation of different lensing fields offers \textit{twice} the number of available measurements as their autocorrelation counterparts. We can define
\begin{equation}
    \xi_{\pm}^{\F\G} \equiv \xi_{\pm}^{\F\rightarrow\G} \cup \xi_{\pm}^{\G\rightarrow\F}
\end{equation}

\noindent as the combination of both $\F\rightarrow\G$ and $\G\rightarrow\F$ correlation measurements. Here, $\cup$ refers to combining both measurements together while accounting for the  algebraic sign of each so as not to cancel to zero. The spin combination for $\F$ and $\G$ is even, so $\xi_{\pm}^{\F\rightarrow\G} = \xi_{\pm}^{\G\rightarrow\F}$. Therefore, for this field combination, $\cup$ is equivalent to addition.

Similarly, one is able to obtain twice the number of galaxy pairs for shear-flexion correlation functions: 
\begin{equation}
    \xi_{\pm}^{\gamma\F} \equiv \xi_{\pm}^{\gamma\rightarrow\F} \cup \xi_{\pm}^{\F\rightarrow\gamma}
\end{equation}

\noindent Here the spin combination is odd and $\cup$ is equivalent to subtraction. The same is true for $\xi_{\pm}^{\G\gamma}$.

\section{Measuring Cosmic Flexion}

\subsection{Practical Estimators for Cosmic Flexion}

Let us consider practical estimators of the correlation functions.  For the case of cosmic shear, one estimates the ellipticity\footnote{Ellipticity is given by $\epsilon = (a-b)/(a+b)\times e^{2i\phi}$, where $a$ and $b$ are the semi-major and semi-minor axes, respectively, and $\phi$ is the position angle.} of a galaxy -- that is, the combination of the effect of shear and an intrinsic ellipticity -- rather than just the shear.  The observable ellipticity $\epsilon_i$ of a galaxy image at angular position $\bm{\vartheta}_i$ is related to the intrinsic ellipticity $\epsilon_i^{\rm s}$ and the shear\footnote{This is actually the reduced shear, $\bm{g}$, which is equal to $\gamma$ in the limit $\kappa \ll 1$.} $\gamma(\bm{\vartheta}_i)$ by \citep{Seitz:1996vf, Kilbinger:2014cea}
\begin{equation}
    \epsilon_i = \epsilon_i^{\rm s} + \gamma(\bm{\vartheta}_i)
\end{equation}

\noindent in the weak lensing regime $\kappa \ll 1$.  In addition to an observed ellipticity, each galaxy may be assigned a weight factor $w_i$ which reflects the measurement uncertainty.  Noisy objects can be down weighted by assigning small values of $w_i$ to them.  We shall assume that the correlation function is to be estimated in bins of some (typically logarithmic) angular width $\Delta\theta$, and we define the function $\Delta_\theta(\phi) = 1$ for angular separations within the bin and zero otherwise.  The standard estimators of the cosmic shear two-point correlation functions are given by \cite{Schneider:2002jd}\footnote{Here, we differ from Ref. \cite{Schneider:2002jd} by having our second summation over only $j>i$ to avoid double counting.}
\begin{equation}
    \hat{\xi}_{\pm}^{\gamma\gamma}(\theta) = \frac{\sum_{i,j>i}w_iw_j(\epsilon_{i1}'\epsilon_{j1}' \pm \epsilon_{i2}'\epsilon_{j2}')\Delta_\theta(ij)}{N_{\rm p}(\theta)}
\end{equation}

\noindent where again, $1$ and $2$ refer to the field components, $(ij)$ is shorthand for  $(|\bm{\vartheta}_i - \bm{\vartheta}_j|)$, and
\begin{equation}\label{eqn:Np}
    N_{\rm p}(\theta) = \sum_{i,j>i}w_iw_j\Delta_\theta(ij),
\end{equation}

\noindent is the effective number of galaxy pairs per angular bin (it is equal to the number of galaxy pairs in the case that all weights are unity), and where the rotated components of the observed ellipticity are defined in analogy to the corresponding shear components in Eq. \eqref{eqn:gam_t_cross}. Ref. \cite{Schneider:2002jd} showed that this is an unbiased estimator of the cosmic shear.

Following similar lines, we can create estimators for generalized spin fields. The observable field $a_i^{\rm o}$ of a galaxy image at angular position ${\bm \vartheta}_i$ is related to the intrinsic field $a_i^{\rm s}$ and the lensing field $a(\bm{\vartheta}_i)$ by
\begin{equation}\label{eqn:a_obs}
a_i^{\rm o} = a_i^{\rm s} + a(\bm{\vartheta}_i)
\end{equation}

\noindent and similarly, $b_j^{\rm o} = b_j^{\rm s} + b(\bm{\vartheta}_j)$.  An estimator for the correlation functions $\xi_{\pm}^{ab}(\theta)$ is then 
\begin{equation}\label{eqn:estimator}
\hat{\xi}_{\pm}^{ab}(\theta) = \frac{\sum_{i,j>i}w_iw_j(a'^{\rm o}_{i1}b'^{\rm o}_{j1} \pm a'^{\rm o}_{i2}b'^{\rm o}_{j2})\Delta_\theta(ij)}{N_{\rm p}(\theta)},
\end{equation}

\noindent where we emphasize that this estimator specifically should be written as $\xi_{\pm}^{a\rightarrow b}$ in the case where $a$ and $b$ are different spin fields. Now, by showing that the expectation value of this estimator is equal to the correlation function, we can prove it is an unbiased estimator of the correlation function. The expectation value of the estimator is obtained by averaging over the intrinsic fields, assumed to be randomly oriented, and an ensemble average of the lensing field. Considering just $\hat{\xi}_{+}^{ab}$, we find
\begin{equation}
\langle a'^{\rm o}_{i1}b'^{\rm o}_{j1} \pm a'^{\rm o}_{i2}b'^{\rm o}_{j2} \rangle = \sigma_{ab}^2\delta_{ij} + \xi_{+}^{ab}(ij)
\end{equation}

\noindent where $\sigma_{ab}^2$ is the dispersion of the intrinsic fields, and we have used the fact that terms of the form
\begin{equation}\label{eqn:rms_intrinsic}
    \langle a_i^{\rm s*}b_{j}^{\rm s}\rangle = \sigma_{ab}^2\delta_{ij} = \sigma_{a}\sigma_{b}\delta_{ij},
\end{equation}

\noindent and that terms of the form $\langle a_i^{\rm s*}b_{j}\rangle=0$, and, by definition, $\langle a'_{i1}b'_{j1} + a'_{i2}b'_{j2} \rangle = \xi_{+}^{ab}(ij)$, from Eq. \eqref{eqn:2PCF_general_plus}. We therefore see that
\begin{equation}
\left\langle \hat{\xi}_{+}^{ab}(\theta) \right\rangle = \xi_{+}^{ab}(\theta)
\end{equation}

\noindent since the term $\sigma_{ab}^2\delta_{ij}\Delta_\theta(ij)$ vanishes for all $i\neq j$, which is the definition of a galaxy pair. This is similarly the case for $\hat{\xi}_{-}^{ab}$.  

The unbiased estimators for the flexion-flexion correlation functions are therefore
\begin{align}
    \label{eqn:xi_FF_est}
    \hat{\xi}_{\pm}^{\F\F}(\theta) &= \frac{\sum_{i,j>i}w_iw_j(\F_{i1}'^{\rm o}\F_{j1}'^{\rm o} \pm \F_{i2}'^{\rm o}\F_{j2}'^{\rm o})\Delta_\theta(ij)}{N_{\rm p}(\theta)} \\
    \label{eqn:xi_GG_est}
    \hat{\xi}_{\pm}^{\G\G}(\theta) &= \frac{\sum_{i,j>i}w_iw_j(\G_{i1}'^{\rm o}\G_{j1}'^{\rm o} \pm \G_{i2}'^{\rm o}\G_{j2}'^{\rm o})\Delta_\theta(ij)}{N_{\rm p}(\theta)} \\
    \label{eqn:xi_FG_est}
    \hat{\xi}_{\pm}^{\F\rightarrow\G}(\theta) &= \frac{\sum_{i,j>i}w_iw_j(\F_{i1}'^{\rm o}\G_{j1}'^{\rm o} \pm \F_{i2}'^{\rm o}\G_{j2}'^{\rm o})\Delta_\theta(ij)}{N_{\rm p}(\theta)} 
\end{align}

\noindent and the (unbiased) estimators for the shear-flexion correlation functions are given by
\begin{align}
    \label{eqn:xi_gammaF_est}
    \hat{\xi}_{\pm}^{\gamma\rightarrow\F}(\theta) &= \frac{\sum_{i,j>i}w_iw_j(\epsilon_{i1}'\F_{j1}'^{\rm o} \pm \epsilon_{i2}'\F_{j2}'^{\rm o})\Delta_\theta(ij)}{N_{\rm p}(\theta)}\\
    \label{eqn:xi_Ggamma_est}
    \hat{\xi}_{\pm}^{\G\rightarrow\gamma}(\theta) &= \frac{\sum_{i,j>i}w_iw_j(\G_{i1}'^{\rm o}\epsilon_{j1}' \pm \G_{i2}'^{\rm o}\epsilon_{j2}')\Delta_\theta(ij)}{N_{\rm p}(\theta)} .
\end{align}

\noindent To this end, we have developed a code capable of computing the flexion and shear correlation functions, known as \texttt{F-SHARP} (Flexion and SHear ARbitrary Point correlations)\footnote{\href{https://github.com/evanjarena/F-SHARP}{https://github.com/evanjarena/F-SHARP}}. This code takes as input the estimated observed flexion and ellipticity components for a set of galaxies, and implements Eqs. \eqref{eqn:xi_FF_est} - \eqref{eqn:xi_Ggamma_est} above to provide correlation function measurements (see for instance Figs. \ref{fg:toy_flexflex} and \ref{fg:toy_shearflex} below).

\subsection{Cosmic Flexion Covariance}

In addition to the two cosmic shear correlation functions, we have described the existence of six flexion-flexion and four shear-flexion correlation functions.  One may wish to calculate covariance matrices for these estimators.  Ref. \cite{Schneider:2002jd} analytically calculated three different covariance matrices for the cosmic shear correlation functions across two different angular bins $\theta_x$ and $\theta_y$: ${\rm Cov}\left(\hat{\xi}_{+}^{\gamma\gamma},\theta_x; \hat{\xi}_{+}^{\gamma\gamma},\theta_y\right)$, ${\rm Cov}\left(\hat{\xi}_{-}^{\gamma\gamma},\theta_x; \hat{\xi}_{-}^{\gamma\gamma},\theta_y\right)$, and ${\rm Cov}\left(\hat{\xi}_{+}^{\gamma\gamma},\theta_x; \hat{\xi}_{-}^{\gamma\gamma},\theta_y\right)$.  Following this approach, we can calculate three covariance matrices for each of the ten additional cosmic flexion and shear-flexion estimators, for a total of 30 additional covariance matrices.  In addition to this, we could calculate the covariance for two different estimators -- for instance, ${\rm Cov}\left(\hat{\xi}_{+}^{\gamma\gamma},\theta_x; \hat{\xi}_{+}^{\gamma\F},\theta_y\right)$.  All told, twelve cosmic weak lensing estimators allow for $12 + 12(12-1)/2 = 78$ possible unique covariance matrices.   

Owing to the large number of covariance matrix permutations, we choose to calculate the most generalized versions:
\[{\rm Cov}\left(\hat{\xi}_{\pm}^{ab},\theta_x; \hat{\xi}_{\pm}^{cd},\theta_y\right) \quad {\rm and} \quad {\rm Cov}\left(\hat{\xi}_{+}^{ab},\theta_x; \hat{\xi}_{-}^{cd},\theta_y\right). \]

\noindent These covariances are derived in Appendix \ref{sec:appendix_B}.

From these covariance matrices, we are able to approximate the autovariance of each estimator -- i.e. the diagonal of ${\rm Cov}\left(\hat{\xi}_{\pm}^{ab},\theta; \hat{\xi}_{\pm}^{ab},\theta\right)$. Under the assumption that the autovariance of the estimators in each bin is dominated by the intrinsic field shape noise, Eq. \eqref{eqn:cov} simply becomes

\begin{equation}
    {\rm Var}\left(\hat{\xi}_{\pm}^{ab}(\theta)\right) \simeq \frac{\sigma_a^2\sigma_b^2}{2\left[N_{\rm p}(\theta)\right]^2}\sum_{i,j>i}w_i^2w_j^2\Delta_{\theta}(ij)
\end{equation}

\noindent where the effective dispersion of the intrinsic field is calculated as
\begin{equation}
    \sigma_a^2 = \frac{\sum_i |a^{\rm o}_i|^2w_i^2}{\sum_i w_i}.
\end{equation}

\noindent Consider the example of the cosmic shear estimators
\begin{equation}
    {\rm Var}\left(\hat{\xi}_{\pm}^{\gamma\gamma}(\theta)\right) \simeq \frac{\sigma_\epsilon^4}{2\left[N_{\rm p}(\theta)\right]^2}\sum_{i,j>i}w_i^2w_j^2\Delta_{\theta}(ij).
\end{equation}
\noindent where the effective dispersion of the intrinsic ellipticity\footnote{In cosmic shear studies, it is often standard practice to measure a dispersion \textbf{per shear component}; however, we choose not to use this formalism.}  
\begin{equation}\label{eqn:sigma_eps}
    \sigma_\epsilon^2 = \frac{\sum_i |\epsilon_i|^2w_i^2}{\sum_i w_i}.
\end{equation}

\noindent In the case of all weights being equal to unity, this expression simplifies to the well known result ${\rm Var}(\hat{\xi}_{\pm}^{\gamma\gamma}(\theta)) \simeq \sigma_\epsilon^4/2N_{\rm p}(\theta)$ given in e.g. Ref. \cite{DES:2020ypx}.

Unlike shear/ellipticity, which is dimensionless, flexion has units of inverse length and is therefore not scale/distance invariant.  The combination of a galaxy's size\footnote{The size of a galaxy, $a$, is not to be confused with the generalized lensing field in previous equations.}, $a = \sqrt{|Q_{11}+Q_{22}|}$ where $Q_{ij}$ are quadrupole image moments, and flexion produces a scale-invariant, dimensionless flexion: $|a\F|$ and $|a\G|$ \citep{Goldberg:2004hh, Fabritius_II_2020}.  We may then define the scatter in intrinsic flexions in the following way:
\begin{align}
    \label{eqn:sigma_aF}
    \sigma_{a\F}^2 &= \frac{\sum_i |a_i\F_i^{\rm o}|^2w_i^2}{\sum_i w_i} \\
    \label{eqn:sigma_aG}
    \sigma_{a\G}^2 &= \frac{\sum_i |a_i\G_i^{\rm o}|^2w_i^2}{\sum_i w_i}.
\end{align}

\noindent The autovariance of the flexion-flexion  estimators is approximated by 
\begin{align}\label{eqn:var_FF}
   {\rm Var}\left(\hat{\xi}_{\pm}^{\F\F}(\theta)\right) &\simeq \frac{\sigma_{a\F}^4}{2\left[N_{\rm p}(\theta)\right]^2}\sum_{i,j>i}\frac{w_i^2w_j^2\Delta_{\theta}(ij)}{a_i^2a_j^2}  \\
   \label{eqn:var_GG}
   {\rm Var}\left(\hat{\xi}_{\pm}^{\G\G}(\theta)\right) &\simeq \frac{\sigma_{a\G}^4}{2\left[N_{\rm p}(\theta)\right]^2}\sum_{i,j>i}\frac{w_i^2w_j^2\Delta_{\theta}(ij)}{a_i^2a_j^2} \\
   \label{eqn:var_FG}
   {\rm Var}\left(\hat{\xi}_{\pm}^{\F\rightarrow\G}(\theta)\right) &\simeq \frac{\sigma_{a\F}^2\sigma_{a\G}^2}{2\left[N_{\rm p}(\theta)\right]^2}\sum_{i,j>i}\frac{w_i^2w_j^2\Delta_{\theta}(ij)}{a_i^2a_j^2},
\end{align}

\noindent and the autovariance of the shear-flexion estimators is
\begin{align}\label{eqn:var_gammaF}
   {\rm Var}\left(\hat{\xi}_{\pm}^{\gamma\rightarrow\F}(\theta)\right) &\simeq \frac{\sigma_{\epsilon}^2\sigma_{a\F}^2}{2\left[N_{\rm p}(\theta)\right]^2}\sum_{i,j>i}\frac{w_i^2w_j^2\Delta_{\theta}(ij)}{a_j^2}  \\
   \label{eqn:var_Ggamma}
   {\rm Var}\left(\hat{\xi}_{\pm}^{\G\rightarrow\gamma}(\theta)\right) &\simeq \frac{\sigma_{a\G}^2\sigma_{\epsilon}^2}{2\left[N_{\rm p}(\theta)\right]^2}\sum_{i,j>i}\frac{w_i^2w_j^2\Delta_{\theta}(ij)}{a_i^2}.
\end{align}

\subsection{Testing Cosmic Flexion with a Gaussian Random Field}

\begin{figure*}[htb!]
    \hspace{-2.5em}
    {\includegraphics[width=0.37\linewidth]{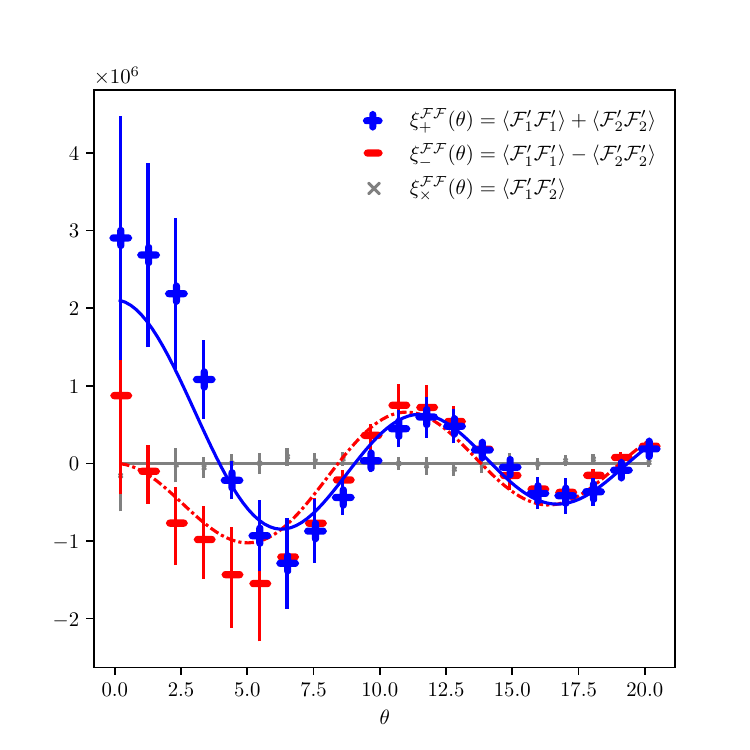}}
    \hspace{-2.5em}
    {\includegraphics[width=0.37\linewidth]{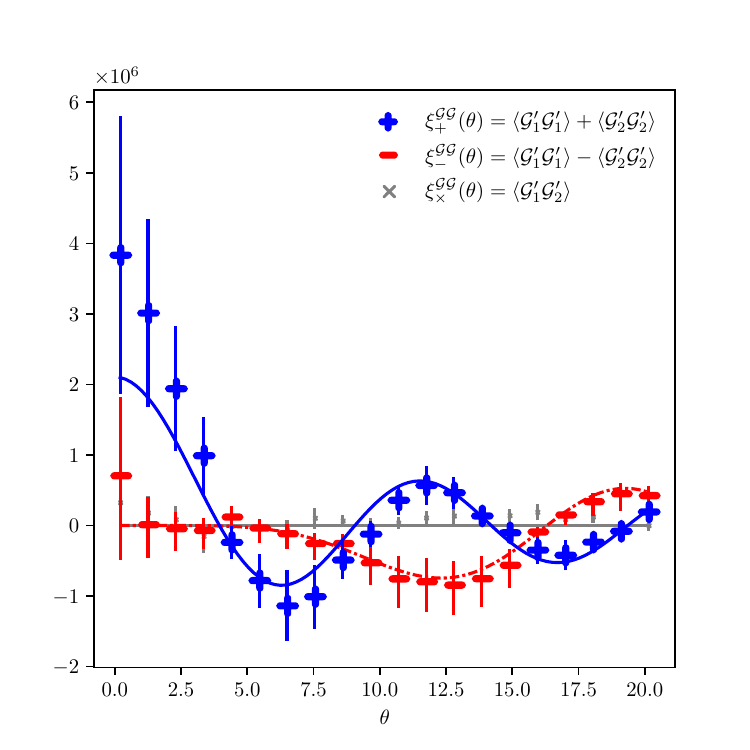}}
    \hspace{-2.5em}
    {\includegraphics[width=0.37\linewidth]{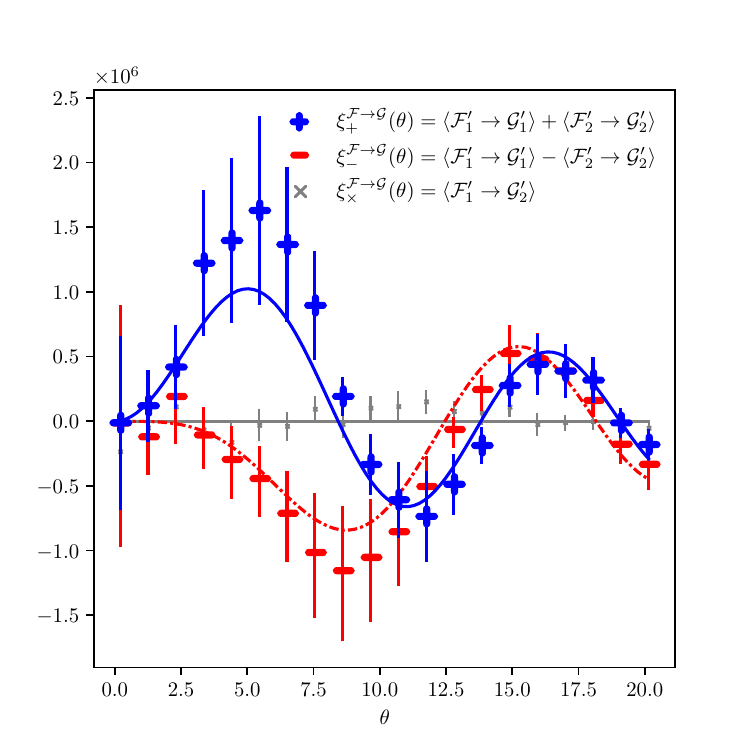}}
    \hspace{-2.5em}
    \caption{Theoretical cosmic flexion-flexion correlation functions $\xi_{\pm}^{\F\F}$, $\xi_{\pm}^{\G\G}$, and $\xi_{\pm}^{\F\rightarrow\G}$ for a delta-function convergence Gaussian random field.  The solid (blue) lines are the `$+$' theoretical correlation functions, and the dash-dotted (red) lines are the `$-$' correlations.  We see that the measurements of the `$+$' and `$-$' correlation functions are consistent with the theoretical curves.  We also see that the so-called ``cross" (`$\times$') correlation functions, which vanish due to parity-symmetry, are consistent with zero. Angular separation, $\theta$, has units of arcseconds, and the flexion-flexion correlation functions have units of [radians]$^{-2}$.}
     \label{fg:toy_flexflex}
\end{figure*}

\begin{figure*}[htb!]
    \hspace{-2.5em}
    {\includegraphics[width=0.37\linewidth]{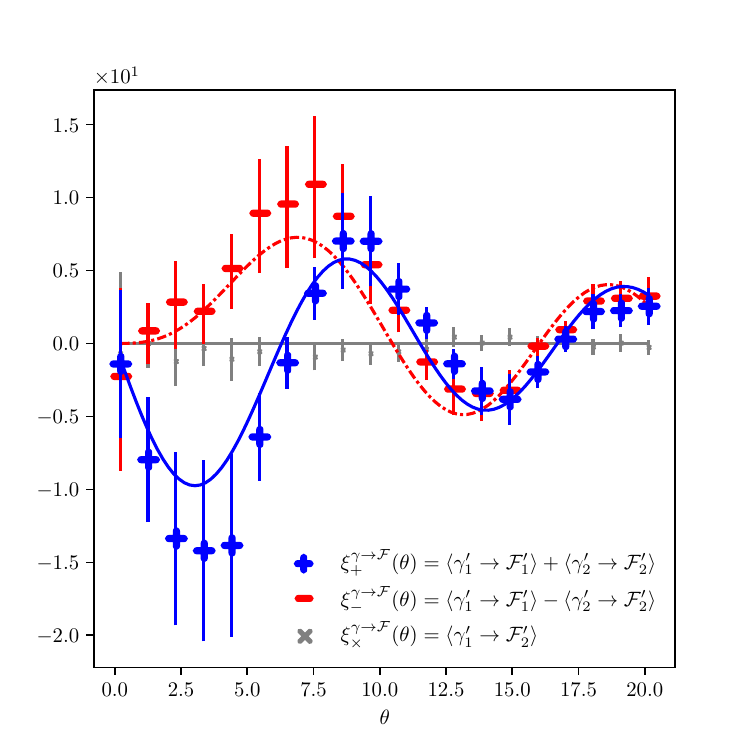}}
    \hspace{-2.5em}
    {\includegraphics[width=0.37\linewidth]{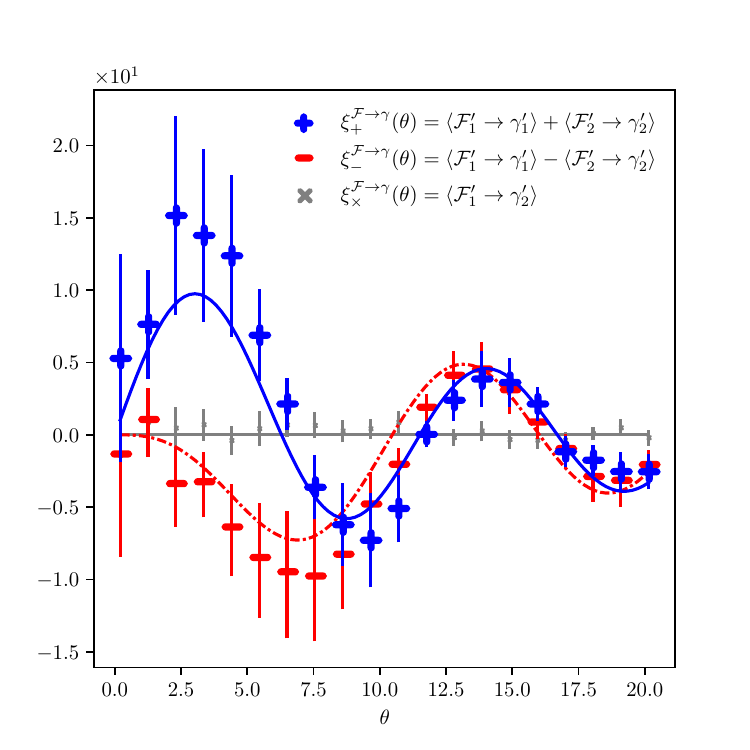}}
    \hspace{-2.5em}
    {\includegraphics[width=0.37\linewidth]{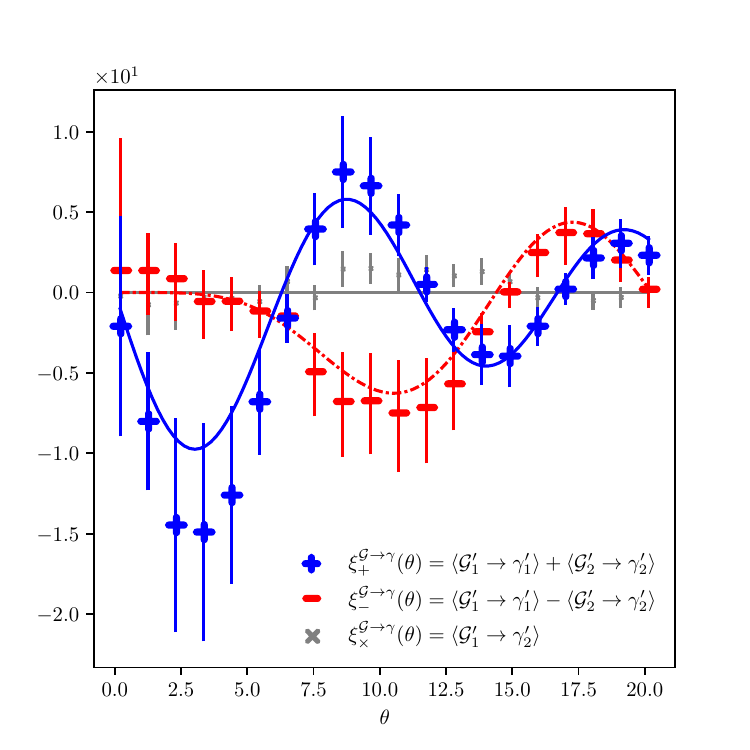}}
    \hspace{-2.5em}
    \caption{Theoretical cosmic shear-flexion correlation functions $\xi_{\pm}^{\gamma\rightarrow\F}$, $\xi_{\pm}^{\F\rightarrow\gamma}$, and $\xi_{\pm}^{\G\rightarrow\gamma}$ for a delta-function convergence Gaussian random field.  The solid (blue) lines are the `$+$' theoretical correlation functions, and the dash-dotted (red) lines are the `$-$' correlations.  We also see that the so-called ``cross" (`$\times$') correlation functions, which vanish due to parity-symmetry, are consistent with zero.  From these plots, we see that $\xi_{\pm}^{\F\rightarrow\gamma} = -\xi_{\pm}^{\gamma\rightarrow\F}$, which verifies Eqs. \eqref{eqn:gammaF_parity_real} and \eqref{eqn:gammaF_parity_Fourier}.  Angular separation, $\theta$ has units of arcseconds, and the shear-flexion correlation functions have units of [radians]$^{-1}$.}
    \label{fg:toy_shearflex}
\end{figure*}

In order to test both our theoretical assumptions and the estimators for the two-point correlation functions, we make use of a simple toy model.  We generate a Gaussian random field for the convergence in Fourier space.  We take this to be a delta-function field, which can be used to obtain the lensing potential via the relation
\begin{equation}
    \tilde{\kappa}(\bm{k}) = -\frac{1}{2}k^2\tilde{\psi}(\bm{k})
\end{equation}

\noindent where we have taken the Fourier transform of $\nabla^2 \psi = 2\kappa$. The shear is obtained in Fourier space by \citep{Kaiser:1994jb, Kilbinger:2014cea}
\begin{align}
    \tilde{\gamma}_1 &= \frac{(k_1^2 - k_2^2)}{k^2}\tilde{\kappa} \nonumber \\
    \tilde{\gamma}_2 &= \frac{2k_1k_2}{k^2}\tilde{\kappa}
\end{align}

\noindent and the flexion via\footnote{These have the opposite sign convention from that in BGRT.  We also correct for a missing factor of two in the $\G$-flexion.} \cite{Bacon:2005qr} 
\begin{align}
    \tilde{\F}_1 &= ik_1\tilde{\kappa} \nonumber \\
    \tilde{\F}_2 &= ik_2\tilde{\kappa} \\
    \tilde{\G}_1 &= \frac{i(k_1^3-3k_1k_2^2)}{k^2}\tilde{\kappa} \nonumber \\
    \tilde{\G}_2 &= \frac{i(3k_1^2k_2-k_2^3)}{k^2}\tilde{\kappa}
\end{align}

\noindent Using these relations, one can create maps of the lensing fields on some patch of sky by using a Fast Fourier Transform. The patch of sky used in this toy problem is approximately $3'\times3'$.  With random sampling, one can obtain measurements of the correlation functions in angular bins.  To do this, \texttt{F-SHARP} makes use of Eqs. \eqref{eqn:xi_FF_est} - \eqref{eqn:xi_Ggamma_est} to compute the estimators of each correlation function.  The noise in this toy problem comes from cosmic variance, so we compute errors of the correlation function measurements over multiple random realizations of the field.  Given the fact that the convergence power spectrum is a delta function, one easily obtains analytical solutions to Eqs. \eqref{eqn:xi_FF_theory} - \eqref{eqn:xi_Ggamma_theory} for the various theoretical correlation functions.  

Figs. \ref{fg:toy_flexflex} and \ref{fg:toy_shearflex} show a comparison of the theoretical versus measured two-point correlation functions. These results demonstrate agreement between our theoretical equations for the correlation functions and the estimators of these correlators coded in \texttt{F-SHARP}.  Most notably, we point out the fact that our results demonstrate $\xi_{\pm}^{\F\rightarrow\gamma} = -\xi_{\pm}^{\gamma\rightarrow\F}$, as posited in our discussion of the non-commutativity of weak lensing fields with odd spin combinations. 

\section{Cosmic Flexion in $\Lambda$CDM for Stage III Lensing Survey}
 
When cosmic flexion was first proposed by BGRT more than a decade ago, there was neither the computational pipeline to compute flexion quickly nor a sufficient dataset for its detection. Now that observations have caught up with theoretical estimates, the time is ripe to measure cosmic flexion, which will give us new insight into cosmic structure
on the arcsecond to arcminute scale.

Stage III lensing surveys such as the Dark Energy Survey (DES), the Kilo-Degree Survey (KiDS) and the Hyper Suprime-Cam Subaru Strategic Program (HSC SSP) are ideal candidates for measuring the
cosmic flexion signal.  As a representative example, in this section we will forecast what could be achieved in measuring flexion correlations with DES.

\begin{figure}[h!]
  \centerline{\includegraphics[width=\linewidth]{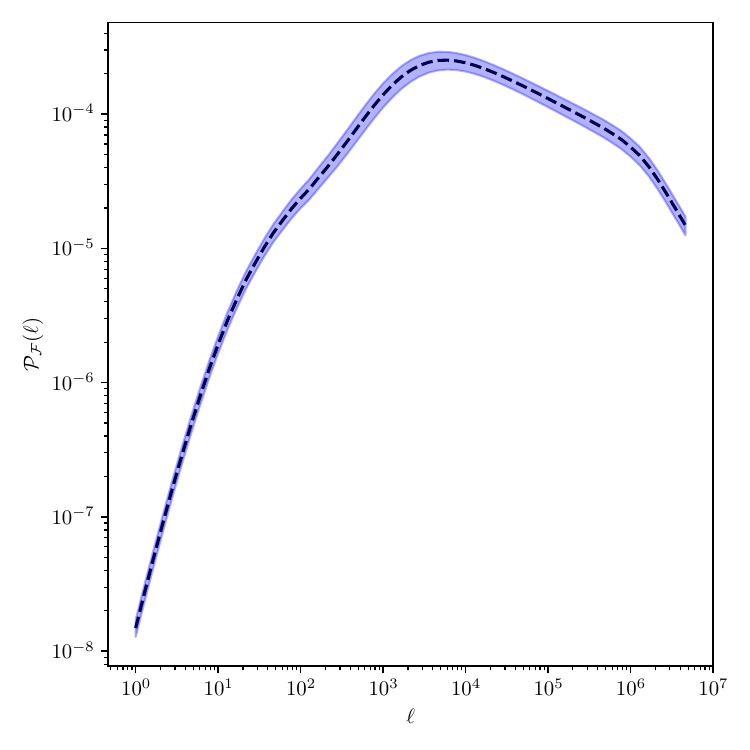}}
  \caption{The cosmic flexion power spectrum expected for the DES Y3 lensing sample using a
    Planck 18 cosmology. The shaded region is the response of the power spectrum to varying
    $\sigma_8$ over ten times the TT,TE,EE+lowE+lensing 68\% interval.  The
    small width of this region is a consequence of the (very) tight
    constraints of the current Planck estimates.  This does not,
    however, include variations of modeling approaches to highly
    nonlinear substructure.}
  \label{fg:flexpower}
\end{figure}

We first calculate the functional form of the cosmic flexion power
spectrum, which is done using \texttt{F-SHARP}. \texttt{F-SHARP} makes use of the Einstein-Boltzmann code \texttt{CLASS} \cite{Blas_2011} to compute the linear matter power spectrum,  which in turn makes use of \textsc{Halofit} \cite{Smith_2003} to compute the nonlinear matter power spectrum.  This assumes a Planck 18 cosmology \cite{Aghanim:2018eyx} using the TT,TE,EE+lowE+lensing constraints.  Next, we make use of the DES (Y3) SOMPZ $n(z)$ source distributions for each redshift bin (as described in Ref. \cite{DES:2021vln}), which are publicly available,\footnote{\href{https://des.ncsa.illinois.edu/releases/y3a2/Y3key-catalogs}{https://des.ncsa.illinois.edu/releases/y3a2/Y3key-catalogs}} combining these in order to estimate the overall source redshift distribution. \texttt{F-SHARP} then makes use of Eqs. \eqref{eqn:q} and \eqref{eqn:P_F} to calculate the flexion power spectrum; this is shown in Fig.~\ref{fg:flexpower}.  Most significantly,
this power spectrum peaks around $\ell\simeq 10^4$ or angular scales on the order of an arcsecond. This should be compared to cosmic shear
measurements, which typically peak on scales $\sim100-1000$ times larger.

\begin{figure}[h!]
  \centerline{\includegraphics[width=\linewidth]{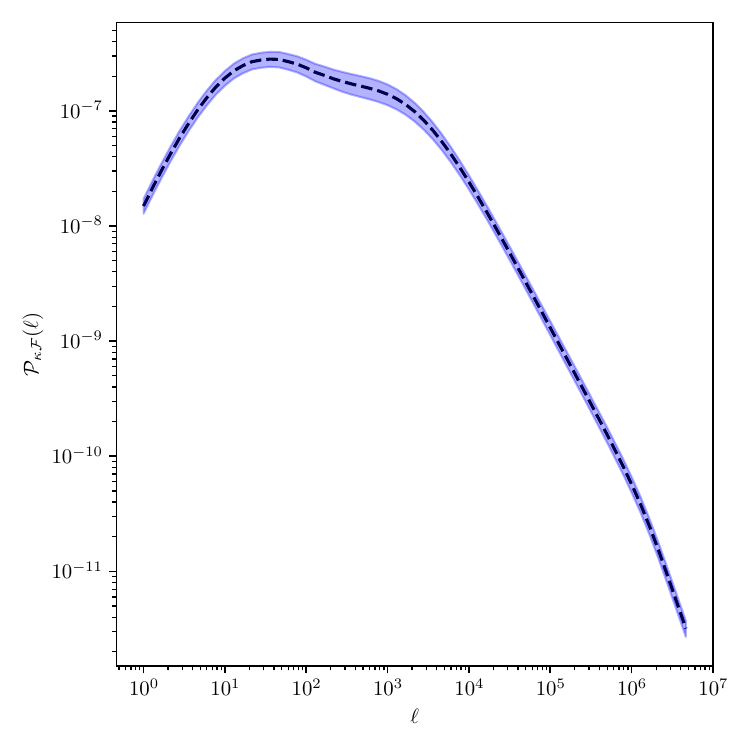}}
  \caption{The cosmic convergence-flexion power spectrum expected for the DES Y3 using a Planck 18 cosmology.  The shaded region is the response of the power spectrum to varying $\sigma_8$ over ten times the TT,TE,EE+lowE+lensing 68\% interval.}
  \label{fg:shearflexpower}
\end{figure}

We also calculate the convergence-flexion power spectrum given by Eq. \eqref{eqn:P_kappaF}; as seen in Fig.~\ref{fg:shearflexpower}, shear-flexion power peaks
  on scales intermediate to that of flexion-flexion and shear-shear. This cross-power bridges the gap between these two probes; since it is (partly)
  measurable in the linear regime ($ \gtrsim 10$ arcminutes), it offers
  the possibility of constraining cosmological parameters and allows for
  systematics checks between cosmic shear and flexion.  
 
 \subsection{Handling Infinities: Renormalization of Cosmic Flexion}
 
 \begin{figure}[htb!]
    \centerline{\includegraphics[width=\linewidth]{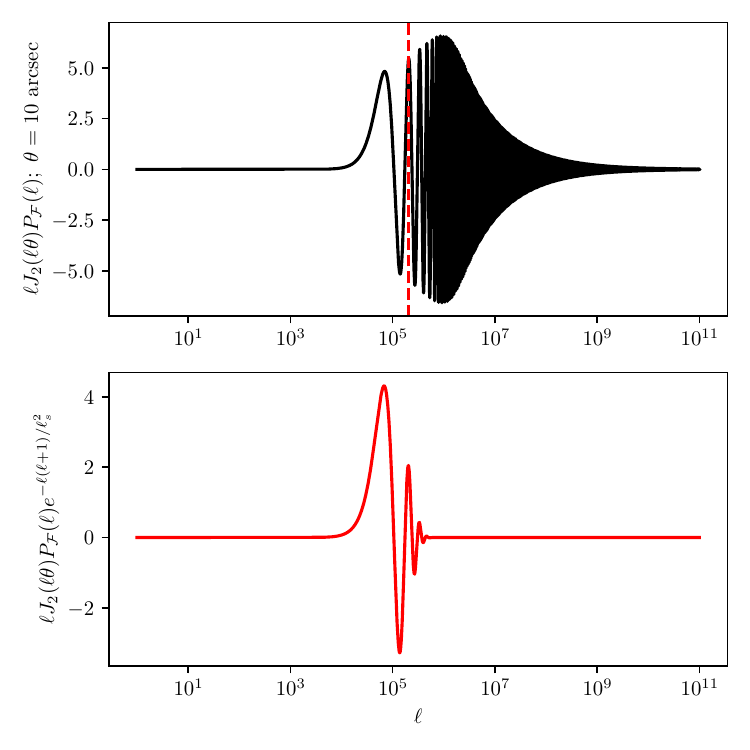}}
    \centerline{\includegraphics[width=\linewidth]{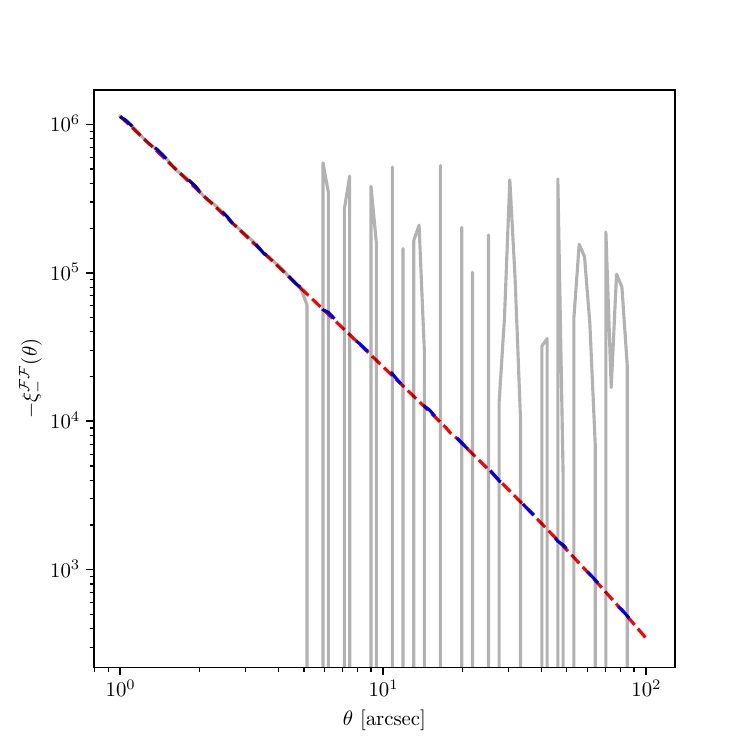}}
    \caption{In the top panel, we show the integrand of the flexion-flexion correlation function $\xi_{-}^{\F\F}$ as a function of $\ell$ for $\theta = 10\,\,{\rm arcseconds}$. We see that after the first peak, there is a rapid oscillation of the cosine envelope.  This ringing makes numerical integration very computationally expensive.  The vertical dashed line is located at the second local maximum, $\ell_s$.  In the middle panel, we show the renormalized integrand, given by Eq. \eqref{eqn:integrand_renorm}.  The bottom panel shows the results of integrating the non-renormalized integrand (solid, gray line) and the normalized integrand (tightly dashed, red line).  Additionally, we show the result of integration via the double-exponential transform of Eq. \eqref{eqn:Hankel_approx} (loosely dashed, blue line).}
 \label{fg:nonconvergence_demo}
\end{figure}
 
When calculating the theoretical correlation functions from the cosmic flexion power spectrum, one encounters integrals that do not converge.  For $\chi \ll \chi_H$, the lensing efficiency scales as $q(\chi) \propto \chi$ (since $a(\chi) \simeq 1$ for $\chi \ll \chi_H$).  Asymptotically, the matter power spectrum follows some power law $\Pow_{\rm NL}(k=\ell/\chi, \chi) \propto k^{-n_s'}$.  Therefore, for low $\chi$ and high $\ell$, the cosmic flexion power spectrum scales as 
 \begin{equation}
     \Pow_\F(\ell) \propto \ell^{2-n_s'} \quad (\rm{asymptotic}).
 \end{equation}
 
\noindent If we examine the integrand of the cosmic flexion-flexion two-point correlation functions, they all have the form   
\begin{equation}
    \frac{d \xi^{\rm flex-flex}}{d\ell} \propto \ell\Pow_\F(\ell) J_n(\ell\theta).
\end{equation}

\noindent Asymptotically, the Bessel functions of the first kind take the form
\begin{equation}
    J_n(x) = \sqrt{\frac{2}{\pi x}}\cos\left(x-(2n+1)\frac{\pi}{4}\right) + \mathcal{O}\left(\frac{1}{x^{3/2}}\right).
\end{equation}

\noindent The integrand then has the asymptotic behavior
\begin{equation}
    \frac{d \xi^{\rm flex-flex}}{d\ell} \propto \ell^{2.5-n'_{s}}\cos\left(\ell\theta -(2n+1)\frac{\pi}{4}\right).
\end{equation}

\noindent Therefore, if $n_s' \leq 2.5$, then these integrals do not converge, because the integral takes the form of a runaway cosine envelope. This is indeed the reality we are faced with if one allows the \textsc{Halofit} routine to compute $P_{\rm NL}$ out to asymptotically large $k$ (or perform a linear extrapolation to arbitrarily large $k$).  However, in Ref. \cite{Widrow:2009}, it is discussed that the matter power spectrum will be proportional to $k^{n_s - 4}$ for arbitrarily large $k$, where $n_s$ is the scaling of the matter power spectrum at low $k$: $P_{\rm NL}(k = \ell/\chi,\chi) \propto k^{n_s}$, where the Planck 18 best fit value for $n_s$ is $\simeq 0.96$ \cite{Aghanim:2018eyx}. 

We therefore propose the following renormalization: compute the matter power spectrum up to some very large $k_{\rm max}$ using the small-scale power spectrum generated by \textsc{Halofit}, and then have the matter power spectrum take the form  $k^{n_s - 4}$ for $k$ beyond that computed by \textsc{Halofit}. This modification of $\Pow_{\rm NL}$ affects the shape of the convergence power spectrum via Eq. \eqref{eqn:P_kappa}, which in turn affects the flexion power spectrum via Eq. \eqref{eqn:P_F}. This allows the cosmic flexion integrals of Eqs. \eqref{eqn:xi_FF_theory} - \eqref{eqn:xi_FG_theory}  to converge.  We note that, since we are only changing the shape of the power spectrum asymptotically, cosmic shear is very insensitive to this renormalization.  For instance, we find that computing $\xi_+^{\gamma\gamma}$ for the non-renormalized and renormalized power spectra are indistinguishable to within one part in $10^7$.  

We next encounter another problem with integration, but this time it is numerical.  For increasingly large $\theta$, these integrals become very difficult and computationally expensive to integrate due to rapid oscillation of the integrand.  Highly oscillatory integrals have been studied extensively in applied mathematics; however, there does not exist a conventional way to numerically handle them \cite{huybrechs_olver_2009}. We therefore offer two possible methods that we find to be in agreement with each other at the percent level.  First, we offer in this paper a novel technique in which we renormalize the integrals given in Eqs. \eqref{eqn:xi_FF_theory} - \eqref{eqn:xi_FG_theory}. Here, we multiply the integrands by a decaying exponential.  These integrands then take the form
\begin{equation}\label{eqn:integrand_renorm}
    \frac{d \xi^{\rm flex-flex}}{d\ell} \propto \ell\Pow_\F(\ell) J_n(\ell\theta)\times e^{-\ell(\ell+1)/\ell_{s}^2}
\end{equation}

\noindent where $\ell_s$ is taken to be the location of the second maximum of the integrand. An alternative method to a second renormalization is an existing technique: a double-exponential variable transformation based on the zeros of the Bessel function of the first kind \citep{Ooura1999, Ogata2005}.  We use the approximation \cite{Szapudi2005} 
\begin{align}\label{eqn:Hankel_approx}
    &\int_0^\infty dx f(x)J_n(x) \nonumber\\
    & \simeq \pi\sum_{k=1}^\infty w_{nk}f\left(\frac{\pi \psi(hr_{nk})}{h}\right)J_n\left(\frac{\pi\psi(hr_{nk})}{h}\right)\psi'(hr_{nk})
\end{align}

\noindent where $r_{nk}$ are the roots of $J_n(x)$ divided by $\pi$, $\psi(t) = t\tanh(\frac{1}{2}\pi\sinh t)$ is the double-exponential transform, $h$ is the step size of the integration, and the weights are $w_{nk} = Y_n(\pi r_{nk})/J_{n+1}(\pi r_{nk})$, where $Y_n$ is the Bessel function of the second kind, order $n$. For our purposes, we take $f(x) \rightarrow \ell\Pow_{\F}(\ell)$ and $J_n(x) \rightarrow J_n(\ell\theta)$.  

A special technique of either renormalization or the double-exponential transform is not necessary for small $\theta$, where the integrand ringing is negligible.  We can therefore test these two approaches by comparing them to the non-renormalized integration at low $\theta$.  These results are shown in Fig. \ref{fg:nonconvergence_demo} for $\xi_{-}^{\F\F}.$ The renormalization integration method is computed using \texttt{F-SHARP} and the double-exponential transform integration method is computed using the public library \texttt{hankel}\footnote{\href{https://github.com/steven-murray/hankel}{https://github.com/steven-murray/hankel}} (see Ref. \cite{Murray2019}). We see that for small $\theta$, where the integrand ringing is minimal and can be easily integrated numerically, all three methods of integration are in agreement. For large $\theta$, where the non-renormalized numerical integration fails, the renormalization and the double-exponential transform allow for efficient and accurate numerical integration.  Again, since both of these techniques agree with each other at the percent level, and are therefore indistinguishable in this context, we can use either.  

\subsection{Forecasts for the Dark Energy Survey}

\begin{figure*}[htb!]
    \centerline{\includegraphics[width=0.75\linewidth]{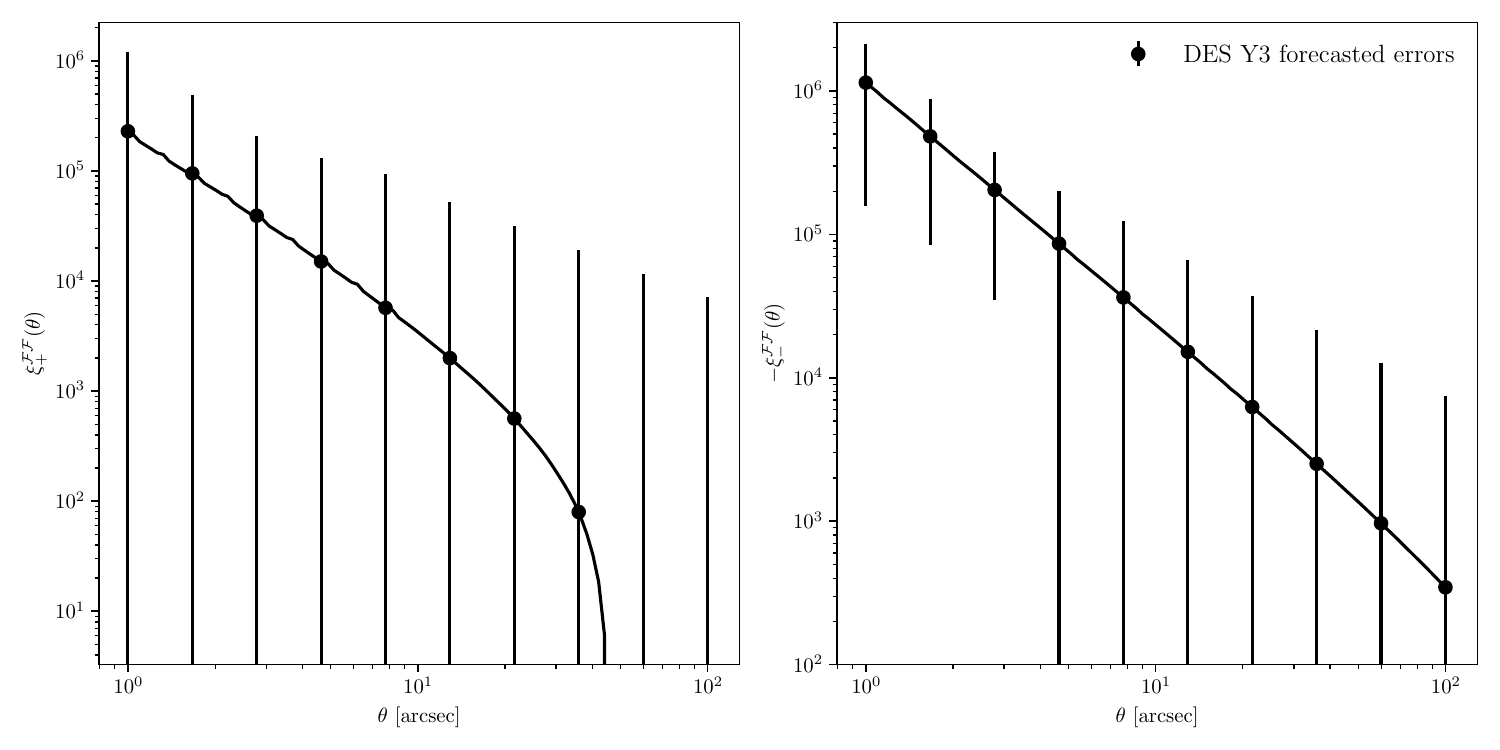}}
    \centerline{\includegraphics[width=0.75\linewidth]{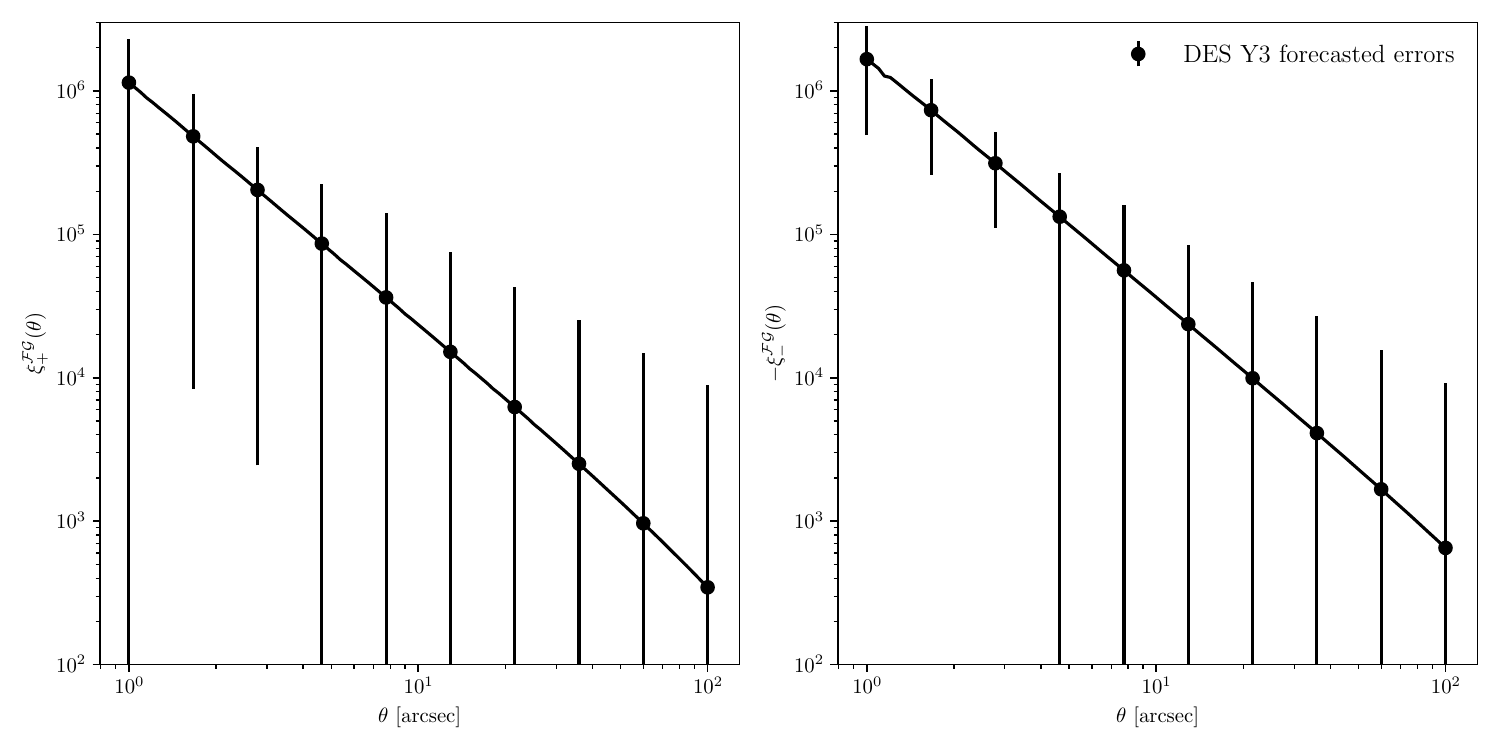}}
    \centerline{\includegraphics[width=0.75\linewidth]{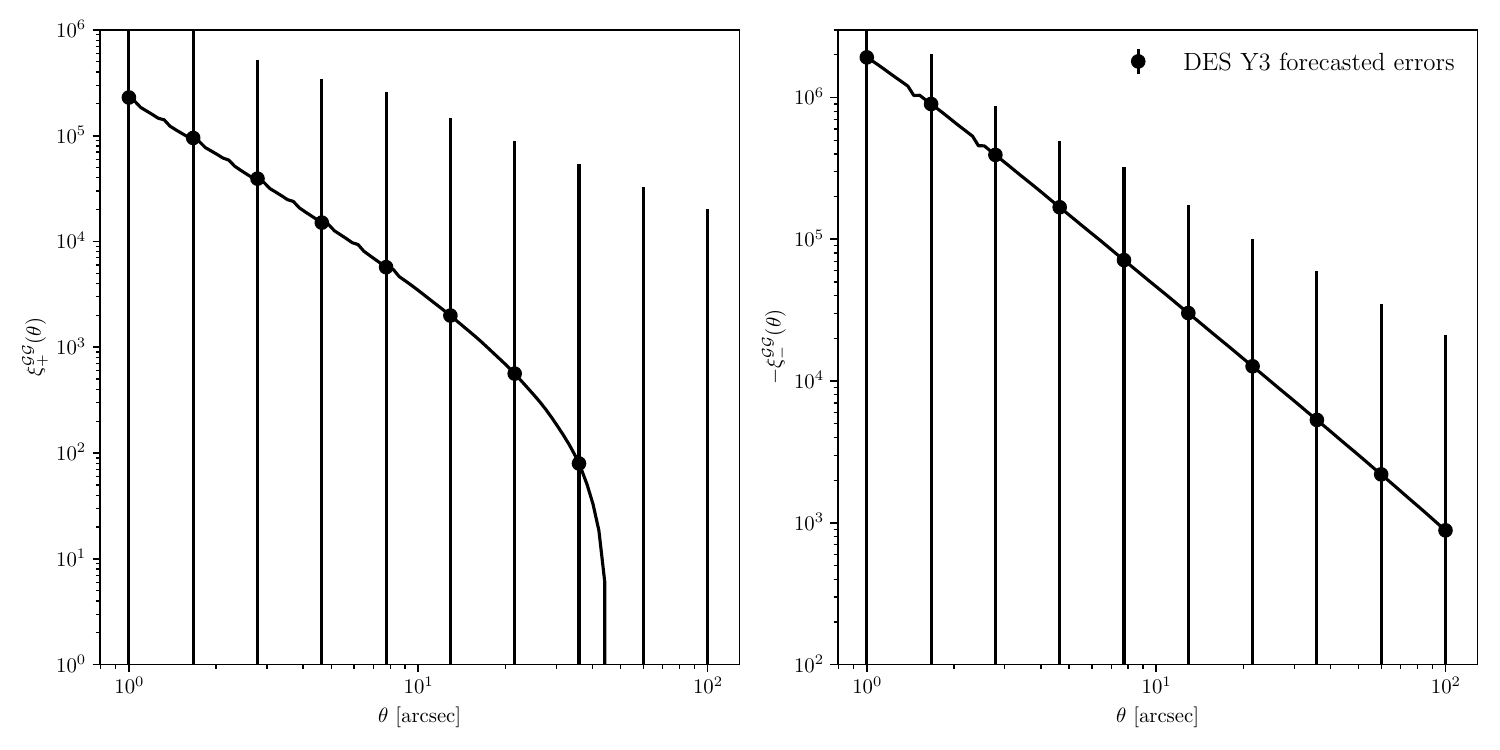}}
    \caption{The theoretical cosmic flexion $\cal F$-$\cal F$
      autocorrelation (top row), $\cal F$-$\cal G$
      cross-correlation (bottom row), and $\G$-$\G$ autocorrelation (bottom row) functions with forecast errors
      for DES Y3.  Note that the data points are equal to the
      theoretical values and do not represent a measurement. Here, we
      anticipate a higher S/N for the $\cal F$-$\cal G$
      cross-correlation than the $\cal F$-$\cal F$ autocorrelation.  Note that $\xi_{\pm}^{\F\G}$ here represents the combined use of both $\xi_{\pm}^{\F\rightarrow\G}$ and $\xi_{\pm}^{\G\rightarrow\F}$.}
 \label{fg:cosmicflex}
\end{figure*}

\begin{figure*}[htb!]
    \centerline{\includegraphics[width=0.75\linewidth]{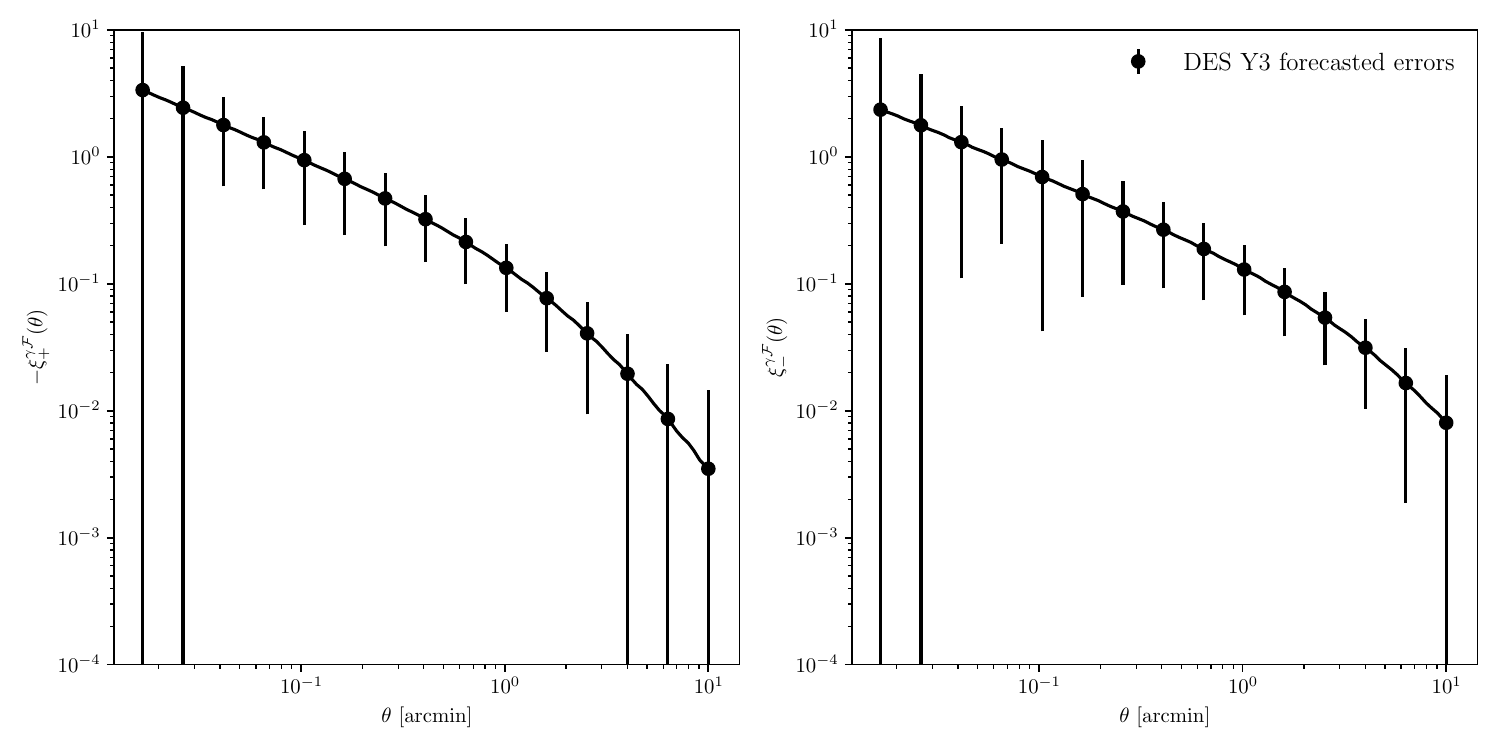}}
    \centerline{\includegraphics[width=0.75\linewidth]{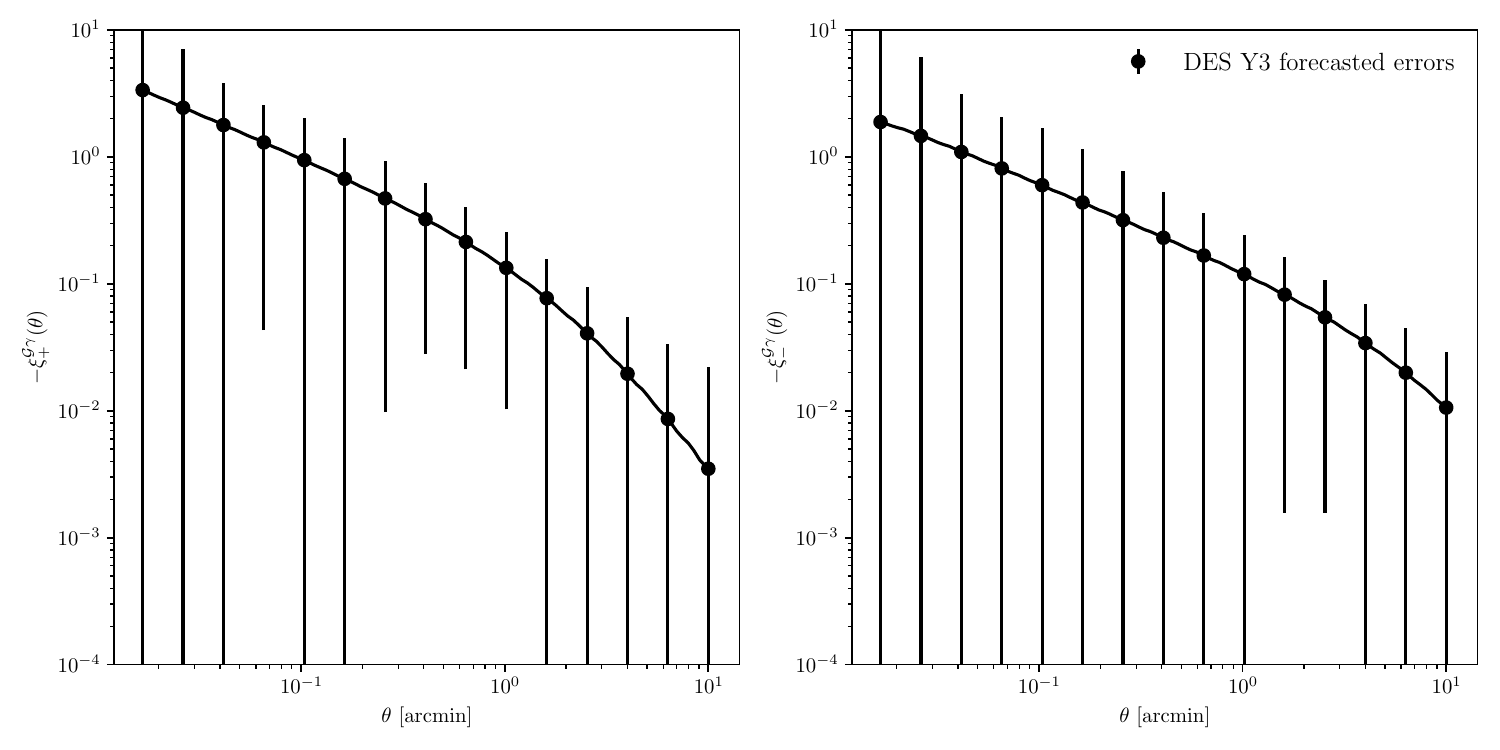}}
    \caption{The theoretical cosmic shear-flexion $\gamma$-$\cal F$
      (top row) and $\G$-$\gamma$ (bottom row) cross-correlation
      functions with forecast errors for DES Y3. Note that the
      data points are equal to the theoretical values and do not
      represent a measurement. Here, we anticipate a much higher S/N
      for shear-flexion than flexion-flexion. Note that $\xi_{\pm}^{\gamma\F}$ here represents the combined use of both $\xi_{\pm}^{\gamma\rightarrow\F}$ and $\xi_{\pm}^{\F\rightarrow\gamma}$ (and similarly for $\xi_{\pm}^{\G\gamma}$).}
 \label{fg:cosmicshearflex}
\end{figure*}

We can preview the expected signal-to-noise of DES flexion correlation functions by measuring  flexion estimators for a small sample of galaxies constituting $\simeq 0.5$ square degree patch of sky, taken from the publicly available DES Shape Catalogue (Y3)\footnote{\href{https://des.ncsa.illinois.edu/releases/y3a2/Y3key-catalogs}{https://des.ncsa.illinois.edu/releases/y3a2/Y3key-catalogs}} (see Ref. \cite{DES:2020ekd}). We retrieve the corresponding galaxy images from the DES Data Management public server.\footnote{\href{https://des.ncsa.illinois.edu}{https://des.ncsa.illinois.edu}} The measurement pipeline for this subsample is as follows: first flexion and ellipticity are measured for each individual galaxy using the code \texttt{Lenser}\footnote{\href{https://github.com/DrexelLenser/Lenser}{https://github.com/DrexelLenser/Lenser}} --  a fast, open source, minimal-dependency Python tool for estimating flexion and shear from real survey data and realistically simulated images (see Ref. \cite{Fabritius_II_2020} for a detailed description). For these forecasts, it is not necessary to measure the correlation functions.  Rather, we are interested in measuring the autovariances of the correlation functions using Eqs. \eqref{eqn:var_FF} - \eqref{eqn:var_Ggamma}.  To do this, we use \texttt{F-SHARP} in order to (i) compute the root mean square noise for the various lensing estimators in the subsample of galaxies (which will remain constant across the entire DES field) using Eqs. \eqref{eqn:sigma_eps} - \eqref{eqn:sigma_aG} and (ii) calculate the number of pairs given by Eq. \eqref{eqn:Np}, which is scaled to the remaining amount of sky in the survey. These are then used in Eqs. \eqref{eqn:var_FF} - \eqref{eqn:var_Ggamma} to calculate predicted errors on the various DES full-survey correlation functions. These forecasts are shown in Figs.~\ref{fg:cosmicflex} and \ref{fg:cosmicshearflex}. The cosmic flexion-flexion correlations are just detectable with the full survey. We immediately see why shear-flexion cross-correlation is a very valuable signal to measure -- it has a much higher S/N than does flexion-flexion. 

Eqs. \eqref{eqn:var_FF} - \eqref{eqn:var_Ggamma} are adequate for calculating errors on the cosmic flexion signals at least in the short term.  
Typically, cosmic shear studies make use of
analytical and/or Gaussian and log-normal simulations to estimate the
covariance matrix of the cosmic shear correlation functions.  This
sub-percent level accuracy of the covariance is necessary in cosmic
shear studies that wish to make likelihood analyses that lead to
constraints on cosmological parameters.  As we do not wish to use the
cosmic flexion results to constrain cosmological parameters, we do not
require this sub-percent level accuracy of the covariance.  
While we have full analytical covariances worked out in Appendix \ref{sec:appendix_B}, they have not yet been tested against, and corrected by, N-body simulations as is done with cosmic shear covariances in DES.  In addition to the fact that we do not require this level of precision on our errors, there also do not currently exist N-body simulations capable of producing weak-lensing maps at a fine enough resolution to study the small-scale structure probed by cosmic flexion.  

\subsection{Discussion}

As we have seen, cosmic flexion peaks at small, nonlinear scales.  These scales are typically discarded in weak lensing studies that seek to only use larger scale information to constrain cosmological parameters. However, the fact that cosmic flexion signals peak at these scales put them in a unique position to constrain the amplitude and shape of this small-scale matter power spectrum, which can lead to a better understanding of the physics at the substructure level.

It is interesting to note that shear-flexion cross-correlation is partly measurable in the large-scale, linear regime ($\gtrsim 10$ arcmin).  One could undertake a study of how the covariance of the shear-flexion cross correlators at these large scales compare with N-body simulations, as is done with cosmic shear.  This could indeed lead to shear-flexion cross-correlation placing constraints on cosmological parameters.  

In the coming decade, cosmologists are preparing to move from the current Stage III experiments such as DES, into the era of Stage IV surveys such as LSST and Euclid. These will offer greater sky coverage than DES, as well as a higher number density of source galaxies \citep{LSSTScience:2009jmu, Refregier:2010ss}.  Higher number density is particularly beneficial for cosmic flexion, as its signal increases with decreasing galaxy separation. As such, Stage IV surveys will allow for much stronger detection of cosmological flexion signals.

Constraints from cosmic flexion in Stage III and IV surveys could allow us to place limits on various models of poorly understood baryonic effects at small, nonlinear scales \cite{Schneider:2019snl}.  It has also been proposed that cosmic flexion could place constraints on primordial non-Gaussianity using Stage IV surveys \cite{Fedeli:2012}.  Finally, we note that cosmic flexion may be an interesting probe of modified gravity; there may exist modified gravity models that deviate from General Relativity+$\Lambda$ on small scales that cannot be detected by cosmic shear or other large-scale cosmological probes \cite{2011arXiv1104.3955C}.

\section{Conclusions}

\noindent In this paper, we have formalized a full theory of cosmic flexion, including flexion-flexion and shear-flexion correlations. We provide a full real- and Fourier-space treatment of the cosmic flexion two-point correlation functions. This includes the previously known signal $\xi_{+}^{\F\F}$ as well as new signals $\xi_{-}^{\F\F}$, $\xi_{+}^{\G\G}$ and $\xi_{-}^{\G\G}$.  We further posit, for the first time, the existence of a cross-correlation between the $\F$- and $\G$-flexions, $\xi_{\pm}^{\F\rightarrow\G}$.  This cross-correlation extends to our treatment of the shear-flexion cross-correlations $\xi_{\pm}^{\gamma\rightarrow\F}$ and $\xi_{\pm}^{\G\rightarrow\gamma}$.  For the first time, we demonstrate that there exists non-commutativity in weak lensing -- and in all odd spin-field combinations in general.  Furthermore, we point out that, provided a given object contains multiple spin fields, generalized two-point cross-correlations yield twice the information of their autocorrelation counterparts.  To our best knowledge, this has never before been demonstrated or considered. 

We have developed measurement techniques for cosmic flexion that consist of estimators and covariances of the cosmic flexion-flexion and flexion-shear two-point correlation functions.  In calculating the covariance of the estimators, we fully work out generalized covariance matrices for all combinations of generalized two-point correlators of any spin-field combination. We have additionally presented the results of testing our theoretical equations by comparing them to measurement of the real-space correlation functions on a Gaussian random field.  To this end, we have developed the code \texttt{F-SHARP} in order to handle the computation of all cosmic flexion correlation functions.  We have presented a renormalization to deal with the nonconvergence of the theoretical correlation function integrals in $\Lambda$CDM cosmologies and also present a technique for efficient numerical computation. 

Finally, we have presented a forecast for cosmic flexion measurements in the Dark Energy Survey Y3 field, indicating that measuring cosmic flexion is currently feasible.   We have also discussed the fact that there is significant cosmological value that cosmic flexion will be able to offer to the upcoming era of cosmology, as we seek to further constrain the large- and small-scale structure of the Universe. 

\begin{acknowledgments}
\noindent This work makes use of publicly available data from the Dark Energy Survey, which can be found at: \href{https://www.darkenergysurvey.org/the-des-project/data-access/}{https://www.darkenergysurvey.org/the-des-project/data-access/}.  E. J. Arena would like to thank Judit Prat for useful discussions regarding DES data.  E. J. Arena would also like to thank the anonymous referee for valuable suggestions regarding the highly oscillatory cosmic flexion integrands.   
\end{acknowledgments}

\onecolumngrid

\appendix

\section{Deriving the Theoretical Two-Point Correlation Functions}\label{sec:appendix_A}

\subsection{Shear-Shear}

It is well known that one can define the convergence power spectrum in the following way under the flat-sky approximation: \cite{Kilbinger:2014cea}
\begin{equation}
    \langle \tilde{\kappa}(\bm{\ell})\tilde{\kappa}^*(\bm{\ell}')\rangle = (2\pi)^2\delta_{\rm D}(\bm{\ell}-\bm{\ell}') \Pow_{\kappa}(\ell),
\end{equation}

\noindent where, due to statistical homogeneity and isotropy, the power spectrum is a function of the modulus of the two-dimensional multipole vector, $\bm{\ell}$ (the Fourier-conjugate of $\bm{\theta}$, in the case of flat-sky approximation). Let us consider the case of cosmic shear.  In Fourier space, the relationship between shear and convergence is given by \cite{Kilbinger:2014cea}
\begin{equation}
    \tilde{\gamma}(\bm{\ell}) = \frac{\ell_1^2 - \ell_2^2 + 2i\ell_1\ell_2}{\ell^2}\tilde{\kappa}(\bm{\ell}) = e^{2i\beta}\tilde{\kappa}(\bm{\ell})
\end{equation}

\noindent where $\beta$ is the polar angle of $\bm{\ell}$, such that $\bm{\ell} = (\ell_1, \ell_2) = (\ell\cos\beta, \ell\sin\beta)$.  We notice that
\begin{equation}
    \langle\tilde{\gamma}(\bm{\ell})\tilde{\gamma}^*(\bm{\ell'}) \rangle = e^{2i(\beta-\beta')}\langle \tilde{\kappa}(\bm{\ell})\tilde{\kappa}^*(\bm{\ell}')\rangle= e^{2i(\beta-\beta')}(2\pi)^2\delta_{\rm D}(\bm{\ell}-\bm{\ell}') \Pow_{\kappa}(\ell).
\end{equation}

\noindent We will now take the Fourier transform of this. By Eq. \eqref{eqn:2PCF_general_plus}, the left-hand side (LHS) of this expression is simply $\xi_{+}^{\gamma\gamma}(\theta)$.  Thus, we have
\begin{equation}
    \xi_{+}^{\gamma\gamma}(\theta) =  \int \frac{d^2\bm{\ell}}{(2\pi)^2} e^{-i\bm{\ell}\cdot\bm{\theta}}\int\frac{d^2\bm{\ell'}}{(2\pi)^2}\langle\tilde{\gamma}(\bm{\ell})\tilde{\gamma}^*(\bm{\ell'}) \rangle = \int\frac{d^2\bm{\ell}}{(2\pi)^2} e^{-i\bm{\ell}\cdot\bm{\theta}}\int\frac{d^2\bm{\ell'}}{(2\pi)^2} e^{2i(\beta-\beta')}(2\pi)^2\delta_{\rm D}(\bm{\ell}-\bm{\ell}') \Pow_{\kappa}(\ell)
\end{equation}

\noindent Upon integration about $\bm{\ell'}$, the delta function picks out $\bm{\ell'} = \bm{\ell}$ and $\beta' = \beta$, leaving us with
\begin{equation}
    \label{eqn:A1}
    \xi_{+}^{\gamma\gamma}(\theta) =  \int \frac{d^2\bm{\ell}}{(2\pi)^2}e^{-i\bm{\ell}\cdot\bm{\theta}} P_{\kappa}(\ell) = \frac{1}{(2\pi)^2} \int_0^\infty d\ell\, \ell P_{\kappa}(\ell) \int_0^{2\pi} d\beta e^{-i\ell\theta\cos\beta}
\end{equation}

\noindent The Bessel integral can be expressed as
\begin{equation}
    J_n(x) = \frac{1}{\pi}\int_0^\pi d\tau \cos(n\pi - x\sin\tau) = \frac{1}{2\pi}\int_{-\pi}^{\pi}d\tau e^{i\left(n(-\frac{\pi}{2}+\tau)+x\cos\tau\right)} = \frac{i^n}{2\pi} \int_0^{2\pi}d\tau e^{-ix\cos\tau}e^{in\tau}
\end{equation}

\noindent where the second expression is obtained from the first via Euler's formula, and the third expression is obtained from the second via the substitution of variables $\tau \longrightarrow \tau + \pi$ and the relation $(e^{-i\pi/2})^n = i^n$. We therefore obtain the useful result
\begin{equation}
    \label{eqn:bessel_integral}
    \int_0^{2\pi}d\beta e^{-i\ell\theta\cos\beta} i^n e^{in\beta} = 2\pi J_n(\ell\theta)
\end{equation}

\noindent Using this expression, Eq. \eqref{eqn:A1} becomes
\begin{equation}
    \xi_{+}^{\gamma\gamma}(\theta) = \int_0^\infty \frac{d\ell\,\ell}{2\pi} P_{\kappa}(\ell)J_0(\ell\theta).
\end{equation}

\noindent The next quantity of interest is $\langle\tilde{\gamma}(\bm{\ell})\tilde{\gamma}(\bm{\ell'}) \rangle$. First, we note that
\begin{equation}
    \langle \tilde{\kappa}(\bm{\ell})\tilde{\kappa}(\bm{\ell}')\rangle = (2\pi)^2\delta_{\rm D}(\bm{\ell}+\bm{\ell}') \Pow_{\kappa}(\ell).
\end{equation}

\noindent Upon integration, the delta function will pick out $\bm{\ell'} = -\bm{\ell}$, necessarily implying that $\tilde{\kappa}(-\bm{\ell}) = \tilde{\kappa}^*(\bm{\ell}).$ Notice that the shear in Fourier space remains unchanged under the transformation 
\begin{equation}
    \bm{\ell} \rightarrow -\bm{\ell} \implies (\ell_1, \ell_2) \rightarrow -(\ell_1,\ell_2) = -(\ell\cos\beta, \ell\sin\beta).
\end{equation}

\noindent and hence $\tilde{\gamma}(-\bm{\ell}) = \tilde{\gamma}(\bm{\ell})$. Understanding that the delta function will be integrated over, we simply note that
\begin{equation}
    \langle\tilde{\gamma}(\bm{\ell})\tilde{\gamma}(-\bm{\ell}) \rangle = e^{4i\beta}(2\pi)^2 \Pow_{\kappa}(\ell).
\end{equation}

\noindent Upon Fourier transformation, making use of Eq. \eqref{eqn:2PCF_general_minus}, this expression yields
\begin{equation}
    \xi_{-}^{\gamma\gamma}(\theta) = \frac{1}{(2\pi)^2} \int_0^\infty d\ell\,\ell P_{\kappa}(\ell) \int_0^{2\pi} d\beta e^{-i\ell\theta\cos\beta}e^{4i\beta} = \int_0^\infty \frac{d\ell\,\ell}{2\pi} P_{\kappa}(\ell) J_4(\ell\theta).
\end{equation}

\subsection{Flexion-Flexion}

In Fourier space, the relationship between $\F$-flexion and convergence is given by \cite{Bacon:2005qr}
\begin{equation}
    \label{eqn:F_Fourier}
    \tilde{\F}(\bm{\ell}) = (i\ell_1 - \ell_2)\tilde{\kappa}(\bm{\ell}) = i\ell e^{i\beta}\tilde{\kappa}(\bm{\ell}).
\end{equation}

\noindent Following the same lines as for cosmic shear, we can obtain a similar expression for $\F$-flexion:
\begin{equation}
   \langle \tilde{\F}(\bm{\ell})\tilde{\F}^*(\bm{\ell}')\rangle = \ell\ell' \langle \tilde{\kappa}(\bm{\ell})\tilde{\kappa}^*(\bm{\ell}')\rangle = (2\pi)^2\delta_{\rm D}(\bm{\ell}-\bm{\ell}') \ell\ell'\Pow_{\kappa}(\ell) 
\end{equation}

\noindent Again, upon integration about $\bm{\ell'}$, the delta function picks out $\bm{\ell'} = \bm{\ell}$ and $\beta' = \beta$.  Noting this, we can simply write
\begin{equation}
   \langle \tilde{\F}(\bm{\ell})\tilde{\F}^*(\bm{\ell})\rangle = (2\pi)^2\Pow_{\F}(\ell) 
\end{equation}

\noindent where we have used Eq. \eqref{eqn:P_F}. Next, we take a Fourier transform of this expression, which leaves us with
\begin{equation}
    \xi_{+}^{\F\F}(\theta) = \frac{1}{(2\pi)^2}\int_0^\infty d\ell\,\ell \Pow_{\F}(\ell) \int_0^{2\pi} d\beta e^{-i\ell\theta\cos\beta} = \int_0^\infty \frac{d\ell\,\ell}{2\pi} \Pow_{\F}(\ell) J_0(\ell\theta).
\end{equation}

\noindent Unlike shear, $\F$-flexion changes sign in Fourier space under the transformation $\bm{\ell} \rightarrow -\bm{\ell}$ (see Eq. \eqref{eqn:F_Fourier}), such that $\tilde{\F}(-\bm{\ell}) = -\tilde{\F}(\bm{\ell})$.  Hence, for the quantity $\langle\tilde{\F}(\bm{\ell})\tilde{\F}(\bm{\ell'}) \rangle$, we have
\begin{equation}
    \langle\tilde{\F}(\bm{\ell})\tilde{\F}(-\bm{\ell}) \rangle = (2\pi)^2\ell^2 e^{2i\beta} \Pow_{\kappa}(\ell) = (2\pi)^2e^{2i\beta}\Pow_\F(\ell).
\end{equation}

\noindent Taking a Fourier transform of this expression leaves us with
\begin{equation}
    \xi_{-}^{\F\F}(\theta) = \frac{1}{(2\pi)^2}\int_0^\infty d\ell\,\ell \Pow_{\F}(\ell) \int_0^{2\pi} d\beta e^{-i\ell\theta\cos\beta}e^{2i\beta} = -\int_0^\infty \frac{d\ell\,\ell}{2\pi} \Pow_{\F}(\ell) J_2(\ell\theta).
\end{equation}


\noindent In Fourier space, the relationship between $\G$-flexion and convergence is given by
\begin{equation}\label{eqn:G_Fourier}
    \tilde{\G}(\bm{\ell}) = \frac{i\ell_1^3 - 3i\ell_1\ell_2^2 - 3\ell_1^2\ell_2 + \ell_2^3}{\ell^2}\tilde{\kappa}(\bm{\ell}) = i\ell e^{3i\beta}\tilde{\kappa}(\bm{\ell}).
\end{equation}

\noindent From here, it is straightforward to derive expressions for $\xi_{\pm}^{\G\G}(\theta)$.  However, there is an additional complication for $\F$-$\G$ cross-correlations.  If we analyze the expression $\langle\tilde{\F}(\bm{\ell})\tilde{\G}^*(\bm{\ell'}) \rangle$, its Fourier transform is not simply given by Eq. \eqref{eqn:2PCF_general_plus}.  Since $\G_1'$ and $\G_2'$ have a sign difference relative to the definitions of $\F_1'$ and $\F_2'$ (see Eqs. \eqref{eqn:FtFr} and \eqref{eqn:GtGr}), the Fourier transform of $\langle\tilde{\F}(\bm{\ell})\tilde{\G}^*(\bm{\ell'}) \rangle$ is actually $-\xi_+^{\F\rightarrow\G}(\theta)$. We have:
\begin{equation}
    \langle\tilde{\F}(\bm{\ell})\tilde{\G}^*(\bm{\ell'})\rangle = \ell\ell'e^{i(\beta-3\beta')}\langle \tilde{\kappa}(\bm{\ell})\tilde{\kappa}^*(\bm{\ell}')\rangle = (2\pi)^2\delta_{\rm D}(\bm{\ell}-\bm{\ell}')\ell\ell'e^{i(\beta-3\beta')}\Pow_{\kappa}(\ell). 
\end{equation}

\noindent Again, upon integration about $\bm{\ell'}$, the delta function picks out $\bm{\ell'} = \bm{\ell}$ and $\beta' = \beta$.  Noting this, we can simply write
\begin{equation}
    \langle\tilde{\F}(\bm{\ell})\tilde{\G}^*(\bm{\ell})\rangle = (2\pi)^2\ell^2e^{-2i\beta}\Pow_\F(\ell).
\end{equation}

\noindent As we stated earlier, the Fourier transform of the LHS is the negative of Eq. \eqref{eqn:2PCF_general_plus}:
\begin{equation}
    -\xi_+^{\F\rightarrow\G}(\theta) = \frac{1}{(2\pi)^2}\int_0^\infty d\ell\,\ell \Pow_{\F}(\ell) \int_0^{2\pi} d\beta e^{-i\ell\theta\cos\beta}e^{-2i\beta} = -\int_0^\infty \frac{d\ell\,\ell}{2\pi} \Pow_{\F}(\ell) J_2(\ell\theta)
\end{equation}

\noindent and therefore
\begin{equation}
    \xi_+^{\F\rightarrow\G}(\theta) = \int_0^\infty \frac{d\ell\,\ell}{2\pi} \Pow_{\F}(\ell) J_2(\ell\theta).
\end{equation}

\noindent $\G$-flexion changes sign in Fourier space under the transformation $\bm{\ell} \rightarrow -\bm{\ell}$ (see Eq. \eqref{eqn:G_Fourier}), such that $\tilde{\G}(-\bm{\ell}) = -\tilde{\G}(\bm{\ell})$.  Hence, for the quantity $\langle\tilde{\F}(\bm{\ell})\tilde{\G}(\bm{\ell'}) \rangle$, we have
\begin{equation}
    \langle\tilde{\F}(\bm{\ell})\tilde{\G}(-\bm{\ell}) \rangle = (2\pi)^2\ell^2 e^{4i\beta} \Pow_{\kappa}(\ell) = (2\pi)^2e^{4i\beta}\Pow_\F(\ell).
\end{equation}

\noindent Taking a Fourier transform of this expression leaves us with
\begin{equation}
    -\xi_{-}^{\F\rightarrow\G}(\theta) = \frac{1}{(2\pi)^2}\int_0^\infty d\ell\,\ell \Pow_{\F}(\ell) \int_0^{2\pi} d\beta e^{-i\ell\theta\cos\beta}e^{4i\beta} = \int_0^\infty \frac{d\ell\,\ell}{2\pi} \Pow_{\F}(\ell) J_4(\ell\theta)
\end{equation}

\noindent and therefore
\begin{equation}
    \xi_-^{\F\rightarrow\G}(\theta) = -\int_0^\infty \frac{d\ell\,\ell}{2\pi} \Pow_{\F}(\ell) J_4(\ell\theta).
\end{equation}

\noindent One can also compute $\langle \tilde{\G}(\bm{\ell})\tilde{\F}^*(\bm{\ell})\rangle$ and $\langle \tilde{\G}(\bm{\ell})\tilde{\F}(\bm{-\ell})\rangle$, which leads to the result $\xi_{\pm}^{\G\rightarrow\F}(\theta) = \xi_{\pm}^{\F\rightarrow\G}(\theta)$. 

\subsection{Shear-Flexion}

Let us consider the correlation $\gamma\rightarrow\F$.  First,
\begin{equation}
    \langle \tilde{\gamma}(\bm{\ell})\tilde{\F}^*(\bm{\ell'})\rangle = -i\ell'e^{2i\beta-i\beta'} \langle \tilde{\kappa}(\bm{\ell})\tilde{\kappa}^*(\bm{\ell}')\rangle = -(2\pi)^2\delta_{\rm D}(\bm{\ell}-\bm{\ell}') i\ell'e^{i(2\beta-\beta')}\Pow_{\kappa}(\ell) 
\end{equation}

\noindent Again, upon integration about $\bm{\ell'}$, the delta function picks out $\bm{\ell'} = \bm{\ell}$ and $\beta' = \beta$.  Noting this, we can simply write
\begin{equation}
    \langle \tilde{\gamma}(\bm{\ell})\tilde{\F}^*(\bm{\ell})\rangle = -(2\pi)^2 i e^{i\beta}\Pow_{\kappa\F}(\ell) 
\end{equation}

\noindent where we have made use of Eq. \eqref{eqn:P_kappaF}. Taking the Fourier transform of this yields
\begin{equation}
    \xi_{+}^{\gamma\rightarrow\F}(\theta) = -\frac{1}{(2\pi)^2}\int_0^\infty d\ell\,\ell \Pow_{\kappa\F}(\ell) \int_0^{2\pi} d\beta e^{-i\ell\theta\cos\beta}ie^{i\beta} = -\int_0^\infty \frac{d\ell\,\ell}{2\pi} \Pow_{\kappa\F}(\ell) J_1(\ell\theta).
\end{equation}

\noindent Next, we consider $\langle \tilde{\gamma}(\bm{\ell})\tilde{\F}(\bm{\ell'})\rangle$.  After dropping the delta function, we have
\begin{equation}
    \langle \tilde{\gamma}(\bm{\ell})\tilde{\F}(\bm{-\ell})\rangle = (2\pi)^2 i e^{3i\beta}\Pow_{\kappa\F}(\ell) 
\end{equation}

\noindent The Fourier transform gives us:
\begin{equation}
    \xi_{-}^{\gamma\rightarrow\F}(\theta) = \frac{1}{(2\pi)^2}\int_0^\infty d\ell\,\ell \Pow_{\kappa\F}(\ell) \int_0^{2\pi} d\beta e^{-i\ell\theta\cos\beta}ie^{3i\beta} = \int_0^\infty \frac{d\ell\,\ell}{2\pi} \Pow_{\kappa\F}(\ell) J_3(\ell\theta).
\end{equation}

\noindent Finally, one can compute $\langle \tilde{\F}(\bm{\ell})\tilde{\gamma}^*(\bm{\ell})\rangle$ and $\langle \tilde{\F}(\bm{\ell})\tilde{\gamma}(\bm{-\ell})\rangle$, which leads to the result $\xi_{\pm}^{\F\rightarrow\gamma}(\theta) = -\xi_{\pm}^{\gamma\rightarrow\F}(\theta)$.  Similarly, one can compute the expressions for the $\G$-$\gamma$ correlations.

\section{Cosmic Flexion Covariances}\label{sec:appendix_B}

Here we calculate the generalized covariance matrices of two generalized two-point correlation function estimators, $\hat{\xi}_{\pm}^{ab}$ and $\hat{\xi}_{\pm}^{cd}$, across two different angular separations $\theta_x$ and $\theta_y$, where $a$, $b$, $c$, and $d$, are four different spin fields. We will not present a closed-form solution here -- rather, we work through the steps necessary to compute individual terms. We follow the analysis of Ref. \cite{Schneider:2002jd}, generalized to arbitrary estimators. 

We begin with the `$++$' and `$--$' covariances:
\begin{equation}\label{eqn:cov}
{\rm Cov}\left(\hat{\xi}_{\pm}^{ab},\theta_x; \hat{\xi}_{\pm}^{cd},\theta_y\right) = \left\langle\left(\hat{\xi}_{\pm}^{ab}(\theta_x)-\xi_{\pm}^{ab}(\theta_x)\right)\left(\hat{\xi}_{\pm}^{cd}(\theta_y)-\xi_{\pm}^{cd}(\theta_y)\right) \right\rangle.
\end{equation}

\noindent The first term we must evaluate is
\begin{equation}\label{eqn:cov_term1}
    \left\langle\hat{\xi}_{\pm}^{ab}(\theta_x)\hat{\xi}_{\pm}^{cd}(\theta_y)\right\rangle = \frac{1}{N_{\rm p}(\theta_x)N_{\rm p}(\theta_y)}\sum_{i,j>i}\sum_{k,\ell>k}w_i w_j w_k w_\ell \left\langle(a'^{\rm o}_{i1}b'^{\rm o}_{j1} \pm a'^{\rm o}_{i2}b'^{\rm o}_{j2})(c'^{\rm o}_{i1}d'^{\rm o}_{j1} \pm c'^{\rm o}_{i2}d'^{\rm o}_{j2}) \right\rangle \Delta_{\theta_x}(ij)\Delta_{\theta_y}(k\ell)
\end{equation}

\noindent where we have used the definition of the estimators given by Eq. \eqref{eqn:estimator}. Now, it is necessary to work in terms of the unrotated coordinate system.  We will demonstrate that we can relate the component lensing two-points $\langle a_{i\alpha}b_{j\beta}\rangle$ to the correlation functions in a simple way, whereas it is not convenient to do so in the rotated formalism.  Notice that we can simply invert Eq. \eqref{eqn:SO(2)} to obtain (up to a factor of ${\rm csgn}\left[(-i)^{s_a}\right]$, which we take to simply be $-1$ here for simplicity)
\begin{equation}\label{eqn:SO(2)_inverse}
    \begin{pmatrix}a_1\\a_2\end{pmatrix} = -\begin{pmatrix}\cos s_a\varphi & -\sin s_a\varphi\\ \sin s_a\varphi & \cos s_a\varphi\end{pmatrix} \begin{pmatrix}a_1'\\a_2'\end{pmatrix}.
\end{equation}

\noindent Using this transformation, we find that
\begin{align}\label{eqn:cov_exp}
& \left\langle(a'^{\rm o}_{i1}b'^{\rm o}_{j1} \pm a'^{\rm o}_{i2}b'^{\rm o}_{j2})(c'^{\rm o}_{i1}d'^{\rm o}_{j1} \pm c'^{\rm o}_{i2}d'^{\rm o}_{j2}) \right\rangle  \nonumber \\
 &= \left\langle\left(a^{\rm o}_{i1}b^{\rm o}_{j1}\cos(s_a\mp s_b)\varphi_{ij} + a^{\rm o}_{i2}b^{\rm o}_{j1}\sin(s_a\mp s_b)\varphi_{ij} \mp a_{i1}^{\rm o}b_{j2}^{\rm o}\sin(s_a\mp s_b)\varphi_{ij} \pm a_{i2}^{\rm o}b_{j2}^{\rm o}\cos(s_a\mp s_b)\varphi_{ij}\right)\right. \nonumber\\
 &\quad\quad\times \left.\left(c^{\rm o}_{k1}d^{\rm o}_{\ell1}\cos(s_c\mp s_d)\varphi_{k\ell} + c^{\rm o}_{k2}d^{\rm o}_{\ell1}\sin(s_c\mp s_d)\varphi_{k\ell} \mp c_{k1}^{\rm o}d_{\ell2}^{\rm o}\sin(s_c\mp s_d)\varphi_{k\ell} \pm c_{k2}^{\rm o}d_{\ell2}^{\rm o}\cos(s_c\mp s_d)\varphi_{k\ell}\right)\right\rangle \nonumber \\
 &= \left\langle a_{i1}^{\rm o}b_{j1}^{\rm o}c_{k1}^{\rm o}d_{\ell1}^{\rm o}\right\rangle\cos(s_a\mp s_b)\varphi_{ij}\cos(s_c\mp s_d)\varphi_{k\ell} + \left\langle a_{i1}^{\rm o}b_{j1}^{\rm o}c_{k2}^{\rm o}d_{\ell1}^{\rm o}\right\rangle\cos(s_a\mp s_b)\varphi_{ij}\sin(s_c\mp s_d)\varphi_{k\ell} \nonumber \\
 &\quad\quad \mp \left\langle a_{i1}^{\rm o}b_{j1}^{\rm o}c_{k1}^{\rm o}d_{\ell2}^{\rm o}\right\rangle\cos(s_a\mp s_b)\varphi_{ij}\sin(s_c\mp s_d)\varphi_{k\ell} \pm \left\langle a_{i1}^{\rm o}b_{j1}^{\rm o}c_{k2}^{\rm o}d_{\ell2}^{\rm o}\right\rangle\cos(s_a\mp s_b)\varphi_{ij}\cos(s_c\mp s_d)\varphi_{k\ell} \nonumber \\
 &\quad\quad + \left\langle a_{i2}^{\rm o}b_{j1}^{\rm o}c_{k1}^{\rm o}d_{\ell1}^{\rm o}\right\rangle\sin(s_a\mp s_b)\varphi_{ij}\cos(s_c\mp s_d)\varphi_{k\ell} + \left\langle a_{i2}^{\rm o}b_{j1}^{\rm o}c_{k2}^{\rm o}d_{\ell1}^{\rm o}\right\rangle\sin(s_a\mp s_b)\varphi_{ij}\sin(s_c\mp s_d)\varphi_{k\ell} \nonumber\\
 &\quad\quad \mp \left\langle a_{i2}^{\rm o}b_{j1}^{\rm o}c_{k1}^{\rm o}d_{\ell2}^{\rm o}\right\rangle\sin(s_a\mp s_b)\varphi_{ij}\sin(s_c\mp s_d)\varphi_{k\ell} \pm \left\langle a_{i2}^{\rm o}b_{j1}^{\rm o}c_{k2}^{\rm o}d_{\ell2}^{\rm o}\right\rangle\sin(s_a\mp s_b)\varphi_{ij}\cos(s_c\mp s_d)\varphi_{k\ell} \nonumber\\
 &\quad\quad \mp \left\langle a_{i1}^{\rm o}b_{j2}^{\rm o}c_{k1}^{\rm o}d_{\ell1}^{\rm o}\right\rangle\sin(s_a\mp s_b)\varphi_{ij}\cos(s_c\mp s_d)\varphi_{k\ell} \mp \left\langle a_{i1}^{\rm o}b_{j2}^{\rm o}c_{k2}^{\rm o}d_{\ell1}^{\rm o}\right\rangle\sin(s_a\mp s_b)\varphi_{ij}\sin(s_c\mp s_d)\varphi_{k\ell} \nonumber \\
 &\quad\quad + \left\langle a_{i1}^{\rm o}b_{j2}^{\rm o}c_{k1}^{\rm o}d_{\ell2}^{\rm o}\right\rangle\sin(s_a\mp s_b)\varphi_{ij}\sin(s_c\mp s_d)\varphi_{k\ell} - \left\langle a_{i1}^{\rm o}b_{j2}^{\rm o}c_{k2}^{\rm o}d_{\ell2}^{\rm o}\right\rangle\sin(s_a\mp s_b)\varphi_{ij}\cos(s_c\mp s_d)\varphi_{k\ell}\nonumber\\
 &\quad\quad \pm \left\langle a_{i2}^{\rm o}b_{j2}^{\rm o}c_{k1}^{\rm o}d_{\ell1}^{\rm o}\right\rangle\cos(s_a\mp s_b)\varphi_{ij}\cos(s_c\mp s_d)\varphi_{k\ell} \pm \left\langle a_{i2}^{\rm o}b_{j2}^{\rm o}c_{k2}^{\rm o}d_{\ell1}^{\rm o}\right\rangle\cos(s_a\mp s_b)\varphi_{ij}\sin(s_c\mp s_d)\varphi_{k\ell} \nonumber \\
 & \quad\quad - \left\langle a_{i2}^{\rm o}b_{j2}^{\rm o}c_{k1}^{\rm o}d_{\ell2}^{\rm o}\right\rangle\cos(s_a\mp s_b)\varphi_{ij}\sin(s_c\mp s_d)\varphi_{k\ell} + \left\langle a_{i2}^{\rm o}b_{j2}^{\rm o}c_{k2}^{\rm o}d_{\ell2}^{\rm o}\right\rangle\cos(s_a\mp s_b)\varphi_{ij}\cos(s_c\mp s_d)\varphi_{k\ell}.
\end{align}

\noindent Next, we need to calculate the four-point correlation functions of the observed fields.  We can generalize these sixteen permutations to $\langle a_{i\alpha}^{\rm o}b_{j\beta}^{\rm o}c_{k\mu}^{\rm o}d_{\ell\nu}^{\rm o}\rangle$, where the Greek letters $\in \{1,2\}$.  Using Eq. \eqref{eqn:a_obs}, we see that 
\begin{equation}
    \left\langle a_{i\alpha}^{\rm o}b_{j\beta}^{\rm o}c_{k\mu}^{\rm o}d_{\ell\nu}^{\rm o}\right\rangle = \left\langle (a_{i\alpha}^{\rm s} + a_{i\alpha})(b_{j\beta}^{\rm s} + b_{j\beta})(c_{k\mu}^{\rm s}+c_{k\mu})(d_{\ell\nu}^{\rm s}+d_{\ell\nu})\right\rangle.
\end{equation}

\noindent Now, using Eq. \eqref{eqn:rms_intrinsic} and noting that since there is no preferred direction on average for intrinsic fields, then
\begin{equation}
    \langle a_{i\alpha}^{\rm s} b_{j\beta}^{\rm s} \rangle = \frac{\sigma_{ab}^2}{2}\delta_{ij}\delta_{\alpha\beta} = \frac{\sigma_{a}\sigma_{b}}{2}\delta_{ij}\delta_{\alpha\beta} 
\end{equation}

\noindent Then, since $\langle a_{i\alpha}^{\rm s}b_{j\beta}^{\rm s}c_{k\mu}d_{\ell\nu} \rangle = \langle a_{i\alpha}^{\rm s}b_{j\beta}^{\rm s} \rangle \langle c_{k\mu}d_{\ell\nu} \rangle = (1/2)\sigma_{ab}^2\delta_{ij}\delta_{\alpha\beta} \langle c_{k\mu}d_{\ell\nu} \rangle $, and further noting that only terms of even power in $a^{\rm s}$ and $a$ survive, we are left with
\begin{align}
    \left\langle a_{i\alpha}^{\rm o}b_{j\beta}^{\rm o}c_{k\mu}^{\rm o}d_{\ell\nu}^{\rm o}\right\rangle &= \left\langle a_{i\alpha}^{\rm s}b_{j\beta}^{\rm s}c_{k\mu}^{\rm s}d_{\ell\nu}^{\rm s}\right\rangle + \frac{\sigma_{ab}^2}{2}\delta_{ij}\delta_{\alpha\beta}\left\langle c_{k\mu}d_{\ell\nu}\right\rangle + \frac{\sigma_{bd}^2}{2}\delta_{j\ell}\delta_{\beta\nu}\left\langle a_{i\alpha}c_{k\mu}\right\rangle +  \frac{\sigma_{bc}^2}{2}\delta_{jk}\delta_{\beta\mu}\left\langle a_{i\alpha}d_{\ell\nu} \right\rangle \nonumber\\
    &\quad\quad + \frac{\sigma_{ad}^2}{2}\delta_{i\ell}\delta_{\alpha\nu}\left\langle b_{j\beta}c_{k\mu}\right\rangle + \frac{\sigma_{ac}^2}{2}\delta_{ik}\delta_{\alpha\mu}\left\langle b_{j\beta}d_{\ell\nu}\right\rangle + \frac{\sigma_{cd}^2}{2}\delta_{k\ell}\delta_{\mu\nu}\left\langle a_{i\alpha}b_{j\beta}\right\rangle + \left\langle a_{i\alpha}b_{j\beta}c_{k\mu}d_{\ell\nu}\right\rangle
\end{align}

\noindent Next, let us consider the four-point functions of the intrinsic and the lensing fields.  We assume that both are Gaussian, so that the four-point function can be written as a sum over products of two-point functions.  Even without the assumption of the intrinsic field being Gaussian, we can note that the four-point function of the intrinsic fields factorizes, since at most two of the indices $i, j, k, l$ are equal.  Therefore, the intrinsic four-point function becomes
\begin{align}
    \left\langle a_{i\alpha}^{\rm s}b_{j\beta}^{\rm s}c_{k\mu}^{\rm s}d_{\ell\nu}^{\rm s}\right\rangle &= \left\langle a_{i\alpha}^{\rm s}b_{j\beta}^{\rm s}\right\rangle \left\langle c_{k\mu}^{\rm s}d_{\ell\nu}^{\rm s}\right\rangle + \left\langle a_{i\alpha}^{\rm s}c_{k\mu}^{\rm s}\right\rangle \left\langle b_{j\beta}^{\rm s}d_{\ell\nu}^{\rm s}\right\rangle + \left\langle a_{i\alpha}^{\rm s}d_{\ell\mu}^{\rm s}\right\rangle \left\langle b_{j\beta}^{\rm s}c_{k\mu}^{\rm s}\right\rangle \nonumber\\
    &= \frac{\sigma_{ab}^2\sigma_{cd}^2}{4}\left(\delta_{ij}\delta_{\alpha\beta}\delta_{k\ell}\delta_{\mu\nu}\right) + \frac{\sigma_{ac}^2\sigma_{bd}^2}{4}\left(\delta_{ik}\delta_{\alpha\mu}\delta_{j\ell}\delta_{\beta\nu}\right) + \frac{\sigma_{ad}^2\sigma_{bc}^2}{4}\left(\delta_{i\ell}\delta_{\alpha\nu}\delta_{jk}\delta_{\beta\nu}\right).
\end{align}

\noindent Before analyzing the lensing four-point, we note that some of the terms in the above expressions can be dropped. The summations in Eq. \eqref{eqn:cov_term1} require $j>i$ and $\ell>k$, so we can simply drop terms that contain $\delta_{ij}$ and/or $\delta_{k\ell}$.  Also in the summation, it is possible to have $k=i$, $k>i$, and $k<i$.\footnote{One may be tempted here to only compute the upper or lower triangle of the covariance matrix, and for example require $k \geq i$ in the summation.  While this is reasonable for e.g. the shear-shear covariance matrices, it is not advisable in general.  This is because the generalized covariance matrices are \textbf{not} symmetric about the diagonal. This can be demonstrated in the case of two angular bins.  The covariance matrix elements would be of the form $\left(ab(\theta_1)cd(\theta_1), ab(\theta_2)cd(\theta_2)\right)$ along the diagonal, and $\left(ab(\theta_1)cd(\theta_2), ab(\theta_2)cd(\theta_1)\right)$ off the diagonal.  These off-diagonal terms are equal only in the case $c=a$ and $d=b$.}  Therefore, terms where $i=\ell$ and $j=k$ individually survive; however, the requirements $j>i$ and $\ell>k$ require that we can never \textit{simultaneously} have $i=\ell$ and $j=k$, so the product $\delta_{i\ell}\delta_{jk}$ vanishes in the sum.  Dropping these terms, and expanding the lensing four-point in the same way as the intrinsic four-point, we are left with 
\begin{align}
    \left\langle a_{i\alpha}^{\rm o}b_{j\beta}^{\rm o}c_{k\mu}^{\rm o}d_{\ell\nu}^{\rm o}\right\rangle &= \frac{\sigma_{ac}^2\sigma_{bd}^2}{4}\left(\delta_{ik}\delta_{\alpha\mu}\delta_{j\ell}\delta_{\beta\nu}\right) +  \frac{\sigma_{bd}^2}{2}\delta_{j\ell}\delta_{\beta\nu}\left\langle a_{i\alpha}c_{k\mu}\right\rangle +  \frac{\sigma_{bc}^2}{2}\delta_{jk}\delta_{\beta\mu}\left\langle a_{i\alpha}d_{\ell\nu} \right\rangle + \frac{\sigma_{ad}^2}{2}\delta_{i\ell}\delta_{\alpha\nu}\left\langle b_{j\beta}c_{k\mu}\right\rangle \nonumber\\
    &\quad\quad +\frac{\sigma_{ac}^2}{2}\delta_{ik}\delta_{\alpha\mu}\left\langle b_{j\beta}d_{\ell\nu}\right\rangle + \left\langle a_{i\alpha}b_{j\beta}\right\rangle \left\langle c_{k\mu}d_{\ell\nu}\right\rangle + \left\langle a_{i\alpha}c_{k\mu}\right\rangle \left\langle b_{j\beta}d_{\ell\nu}\right\rangle + \left\langle a_{i\alpha}d_{\ell\mu}\right\rangle \left\langle b_{j\beta}c_{k\mu}\right\rangle.
\end{align}

\noindent The next step is to express these two-point functions in terms of the correlation functions.  Using Eqs. \eqref{eqn:2PCF_general_plus}, \eqref{eqn:2PCF_general_minus}, and \eqref{eqn:SO(2)_inverse}, and noting that terms of the form $\langle a'_{i1}b'_{j2}\rangle$ vanish due to parity in the rotated coordinate system, we find that 
\begin{align}\label{eqn:unrot_to_rot}
    \langle a_{i1}b_{j1} \rangle &= \frac{1}{2}\left\{\xi_{+}^{ab}(ij)\cos\left[(s_a-s_b)\varphi_{ij}\right] + \xi_{-}^{ab}(ij)\cos\left[(s_a+s_b)\varphi_{ij}\right] \right\} \nonumber \\
    \langle a_{i2}b_{j2} \rangle &= \frac{1}{2}\left\{\xi_{+}^{ab}(ij)\cos\left[(s_a-s_b)\varphi_{ij}\right] - \xi_{-}^{ab}(ij)\cos\left[(s_a+s_b)\varphi_{ij}\right] \right\} \\
    \langle a_{i1}b_{j2} \rangle &= \frac{1}{2}\left\{-\xi_{+}^{ab}(ij)\sin\left[(s_a-s_b)\varphi_{ij}\right] + \xi_{-}^{ab}(ij)\sin\left[(s_a+s_b)\varphi_{ij}\right] \right\}. \nonumber
\end{align}

\noindent The second and third covariance terms we must evaluate are
\begin{align}\label{eqn:cov_term2}
    \left\langle{\xi}_{\pm}^{ab}(\theta_x)\hat{\xi}_{\pm}^{cd}(\theta_y)\right\rangle &= \frac{1}{N_{\rm p}(\theta_y)}\sum_{k,\ell>k}w_k w_\ell \left\langle(a'_{i1}b'_{j1} \pm a'_{i2}b'_{j2})(c'^{\rm o}_{i1}d'^{\rm o}_{j1} \pm c'^{\rm o}_{i2}d'^{\rm o}_{j2}) \right\rangle \Delta_{\theta_y}(k\ell) \nonumber \\
    \left\langle\hat{\xi}_{\pm}^{ab}(\theta_x)\xi_{\pm}^{cd}(\theta_y)\right\rangle &= \frac{1}{N_{\rm p}(\theta_x)}\sum_{i,j>i}w_i w_j \left\langle(a'^{\rm o}_{i1}b'^{\rm o}_{j1} \pm a'^{\rm o}_{i2}b'^{\rm o}_{j2})(c'_{i1}d'_{j1} \pm c'_{i2}d'_{j2}) \right\rangle \Delta_{\theta_x}(ij).
\end{align}

\noindent Consider expanding the expectation value in these terms.  They are simply Eq. \eqref{eqn:cov_exp} with the replacements $(a^{o} \rightarrow a, b^{\rm o} \rightarrow b)$ and $(c^{o} \rightarrow c, d^{\rm o} \rightarrow d)$, respectively.  For each of these covariance terms, only the lensing four-point functions survive in the generalized terms.  This is because we drop terms that are not even in $a^{\rm s}$, and we also drop terms where $\ell = k$ and $j = i$, which appear in the second and third covariance terms, respectively. After ignoring these terms, we are left with
\begin{equation}
    \left\langle a_{i\alpha}b_{j\beta}c_{k\mu}^{\rm o}d_{\ell\nu}^{\rm o}\right\rangle = \left\langle a_{i\alpha}^{\rm o}b_{j\beta}^{\rm o}c_{k\mu}d_{\ell\nu}\right\rangle = \left\langle a_{i\alpha}b_{j\beta}\right\rangle \left\langle c_{k\mu}d_{\ell\nu}\right\rangle + \left\langle a_{i\alpha}c_{k\mu}\right\rangle \left\langle b_{j\beta}d_{\ell\nu}\right\rangle + \left\langle a_{i\alpha}d_{\ell\mu}\right\rangle \left\langle b_{j\beta}c_{k\mu}\right\rangle.
\end{equation}

\noindent Finally, the fourth covariance term is 
\begin{equation}
\left\langle{\xi}_{\pm}^{ab}(\theta_x){\xi}_{\pm}^{cd}(\theta_y)\right\rangle = \left\langle(a'_{i1}b'_{j1} \pm a'_{i2}b'_{j2})(c'_{i1}d'_{j1} \pm c'_{i2}d'_{j2}) \right\rangle.
\end{equation}

\noindent Expanding this term gives us Eq. \eqref{eqn:cov_exp} with the replacement $(a^{o} \rightarrow a, b^{\rm o} \rightarrow b, c^{o} \rightarrow c, d^{\rm o} \rightarrow d)$. Again, the only generalized term that survives is the lensing four-point function. 

Finally, there also exists the `$+-$' covariance:
\begin{equation}
{\rm Cov}\left(\hat{\xi}_{+}^{ab},\theta_x; \hat{\xi}_{-}^{cd},\theta_y\right) = \left\langle\left(\hat{\xi}_{+}^{ab}(\theta_x)-\xi_{+}^{ab}(\theta_x)\right)\left(\hat{\xi}_{-}^{cd}(\theta_y)-\xi_{-}^{cd}(\theta_y)\right) \right\rangle.
\end{equation}

\noindent where the first covariance term is
\begin{equation}
    \left\langle\hat{\xi}_{\pm}^{ab}(\theta_x)\hat{\xi}_{\pm}^{cd}(\theta_y)\right\rangle = \frac{1}{N_{\rm p}(\theta_x)N_{\rm p}(\theta_y)}\sum_{i,j>i}\sum_{k,\ell>k}w_i w_j w_k w_\ell \left\langle(a'^{\rm o}_{i1}b'^{\rm o}_{j1} + a'^{\rm o}_{i2}b'^{\rm o}_{j2})(c'^{\rm o}_{i1}d'^{\rm o}_{j1} - c'^{\rm o}_{i2}d'^{\rm o}_{j2}) \right\rangle \Delta_{\theta_x}(ij)\Delta_{\theta_y}(k\ell).
\end{equation}

\noindent Using the transformation in Eq. \eqref{eqn:SO(2)_inverse}, we find that
\begin{align}
& \left\langle(a'^{\rm o}_{i1}b'^{\rm o}_{j1} + a'^{\rm o}_{i2}b'^{\rm o}_{j2})(c'^{\rm o}_{i1}d'^{\rm o}_{j1} - c'^{\rm o}_{i2}d'^{\rm o}_{j2}) \right\rangle  \nonumber \\
 &= \left\langle\left(a^{\rm o}_{i1}b^{\rm o}_{j1}\cos(s_a- s_b)\varphi_{ij} + a^{\rm o}_{i2}b^{\rm o}_{j1}\sin(s_a- s_b)\varphi_{ij} - a_{i1}^{\rm o}b_{j2}^{\rm o}\sin(s_a- s_b)\varphi_{ij} + a_{i2}^{\rm o}b_{j2}^{\rm o}\cos(s_a- s_b)\varphi_{ij}\right)\right. \nonumber\\
 &\quad\quad\times \left.\left(c^{\rm o}_{k1}d^{\rm o}_{\ell1}\cos(s_c+ s_d)\varphi_{k\ell} + c^{\rm o}_{k2}d^{\rm o}_{\ell1}\sin(s_c+ s_d)\varphi_{k\ell} + c_{k1}^{\rm o}d_{\ell2}^{\rm o}\sin(s_c+ s_d)\varphi_{k\ell} - c_{k2}^{\rm o}d_{\ell2}^{\rm o}\cos(s_c+ s_d)\varphi_{k\ell}\right)\right\rangle
\end{align}

\noindent We note that the remaining calculation follows the same lines as before.

\twocolumngrid

\bibliography{bibliography}

\end{document}